%% file: SOM.tex
\documentclass[12pt, twocolumn, numberedappendix, twocolappendix]{openjournal}

\usepackage{xcolor}
\usepackage{textgreek}
\usepackage[utf8]{inputenc}
\usepackage[english]{babel}
\usepackage{hyperref}
\hypersetup{
    colorlinks=true,
    linkcolor=blue,
    filecolor=blue,      
    urlcolor=blue,
    citecolor=blue,
}
\usepackage{color,colortbl}
\usepackage{tensind}
\tensordelimiter{?}
\DeclareGraphicsExtensions{.bmp,.png,.jpg,.pdf}
\usepackage{verbatim}
\usepackage[normalem]{ulem}
\usepackage{soul}

\usepackage{newtxtext,newtxmath}
\usepackage[T1]{fontenc}
\usepackage{ae,aecompl}
\usepackage{multirow}
\usepackage{pifont}
\usepackage{enumitem, footmisc}
\usepackage{cprotect}
\usepackage{booktabs}
\newcolumntype{L}{>{$}l<{$}}
\newcolumntype{C}{>{$}c<{$}}
\newcolumntype{R}{>{$}r<{$}}

\usepackage{savesym}
\savesymbol{tablenum}
\usepackage{siunitx}
\restoresymbol{SIX}{tablenum}

\newcommand{\p}{\scalebox{0.6}{+}}
\newcommand{\pp}{\scalebox{0.6}{++}}
\newcommand{\edit}[1]{\textcolor{black}{#1}}

\renewcommand{\d}[1]{\ensuremath{\operatorname{d}\!{#1}}}

\usepackage{graphicx}	
\usepackage{amsmath}	

\graphicspath{ {./figs/} }

\begin{document}
\journalinfo{The Open Journal of Astrophysics}
\submitted{submitted XXX; accepted YYY}

\title{Catalog-based Detection of Unrecognized Blends in Deep Optical Ground Based Imaging}

\author{
Shuang Liang$^{\star1,2}$,
Prakruth Adari$^{\dagger2}$,
Anja von der Linden$^{\ddagger2}$,
The LSST Dark Energy Science Collaboration}

\thanks{$^\star$ E-mail: \href{mailto:sliang92@stanford.edu}{sliang92@stanford.edu}}
\thanks{$\dagger$ E-mail: \href{mailto:prakruth.adari@stonybrook.edu}{prakruth.adari@stonybrook.edu}}
\thanks{$\ddagger$ E-mail: \href{mailto:anja.vonderlinden@stonybrook.edu}{anja.vonderlinden@stonybrook.edu}}


\affiliation{$^{1}$Kavli Institute for Particle Astrophysics and Cosmology, Department of Physics, Stanford University, Stanford, CA 94305}
\affiliation{$^{2}$Department of Physics and Astronomy, Stony Brook University, Stony Brook, NY 11794}

\date{\today}

\begin{abstract}
In deep, ground-based imaging, about 15\,\%-30\,\% of object detections are expected to correspond to two or more true objects - these are called ``unrecognized blends''.
We use Machine Learning algorithms to detect unrecognized blends in deep ground-based photometry using only catalog-level information: colors, magnitude, and size.
We compare the performance of Self Organizing Map, Random Forest, k-Nearest Neighbors, and Anomaly Detection algorithms. We test all algorithms on 9-band (\textit{uBVr$i^{\p}z^{\pp}$YJH}) and 1-size (flux\_radius in \textit{i}-band) measurements of the ground-based COSMOS catalog, and use COSMOS HST data as the truth for unrecognized blend. We find that 17\,\% of objects in the ground-based COSMOS catalog are unrecognized blends. We show that some unrecognized blends can be identified as such using only catalog-level information; but not all blends can be easily identified. Nonetheless, our methods can be used to improve sample purity, and can identify approximately 30\,\% to 80\,\% of unrecognized blends while rejecting 10\,\% to 50\,\% of all detected galaxies (blended or unblended). The results are similar when only optical bands (\textit{uBVr$i^{\p}z^{\pp}$}) and the size information is available. We also investigate the ability of these algorithms to remove photo-z outliers (identified with spectroscopic redshifts), and find that algorithms targeting color outliers perform better than algorithms targeting unrecognized blends.    
Our method can offer a cleaner galaxy sample with lower blending rates for future cosmological surveys such as the Legacy Survey of Space and Time (LSST), and can potentially improve the accuracy on cosmological parameter constraints at a moderate cost of precision.
\end{abstract}


\maketitle


\input{1.0.tex}

\input{2.1.tex}

\input{2.2.tex}
\input{2.3.tex}
\input{2.4.tex}
\input{2.5.tex}

\input{2.6.tex}

\input{3.0.tex}
\input{3.1.tex}
\input{3.2.tex}
\input{3.3.tex}
\input{3.4.tex}
\input{3.5.tex}

\input{4.0.tex}

\input{4.1.tex}

\input{4.2.tex}
\input{4.3.tex}

\input{4.4.tex}
\input{4.5.tex}

\input{5.2.tex}

\input{summary.tex}

\input{ack.tex}

\input{data.tex}


\newpage
\bibliographystyle{mnras}
\bibliography{SOM}

\begin{appendix}
    \input{app.tex}

\end{appendix}

\end{document}

%% file: 1.0.tex
\section{Introduction}
Modern optical cosmology surveys face great challenges due to increasing survey depth and sample size. For example, the Kilo-Degree Survey \citep[KiDS;][]{kids} measures $\sim 10^8$ sources at a $5\sigma$ depth of $r\sim25$; the Dark Energy Survey \citep[DES;][]{des} measures $\sim 3\times10^8$ sources at a $10\sigma$ depth of $r\sim24.3$, and the Hyper Suprime-Cam Subaru Strategic Program Survey \citep[HSC-SSP;][]{hsc} measures $\sim 10^8$ galaxies at a $5\sigma$ depth with $r\sim26$. The Legacy Survey of Space and Time \citep[LSST;][]{lsst} to be conducted at Vera C. Rubin Observatory will observe $\sim 10^{10}$ galaxies at a $5\sigma$ limiting magnitude of $r\sim27$ (in 10 years). 

When the number density of sources increases, there is a greater chance for them to overlap in projection, causing a ``blend''. 
This is especially a concern to ground-based surveys, where the sources are smeared by the atmospheric point-spread function (PSF), which significantly increases the blend rate.
The Dark Energy Science Collaboration (DESC) white paper \citep{lsst_white} estimates that, with the expected galaxy density at a limiting magnitude of $i\sim 26$, almost all (99.5\% of) randomly placed circular apertures with 4 arcsec radius are overlapping.
The fraction of overlaps is 74\% for 2 arcsec apertures and 24\% for 1 arcsec apertures.
At HSC depth (similar to LSST), \citet{hsc_pipe} show that 58\% of objects were deblended, i.e., have multiple peaks in surface brightness in one segmentation. 
\citet{sanchez} estimate that at 10-year depth for LSST, 62\% of galaxies will have at least 1\% flux contributed from neighbouring sources. 

\begin{figure*}
    \centering
    \includegraphics[width=.8\linewidth]{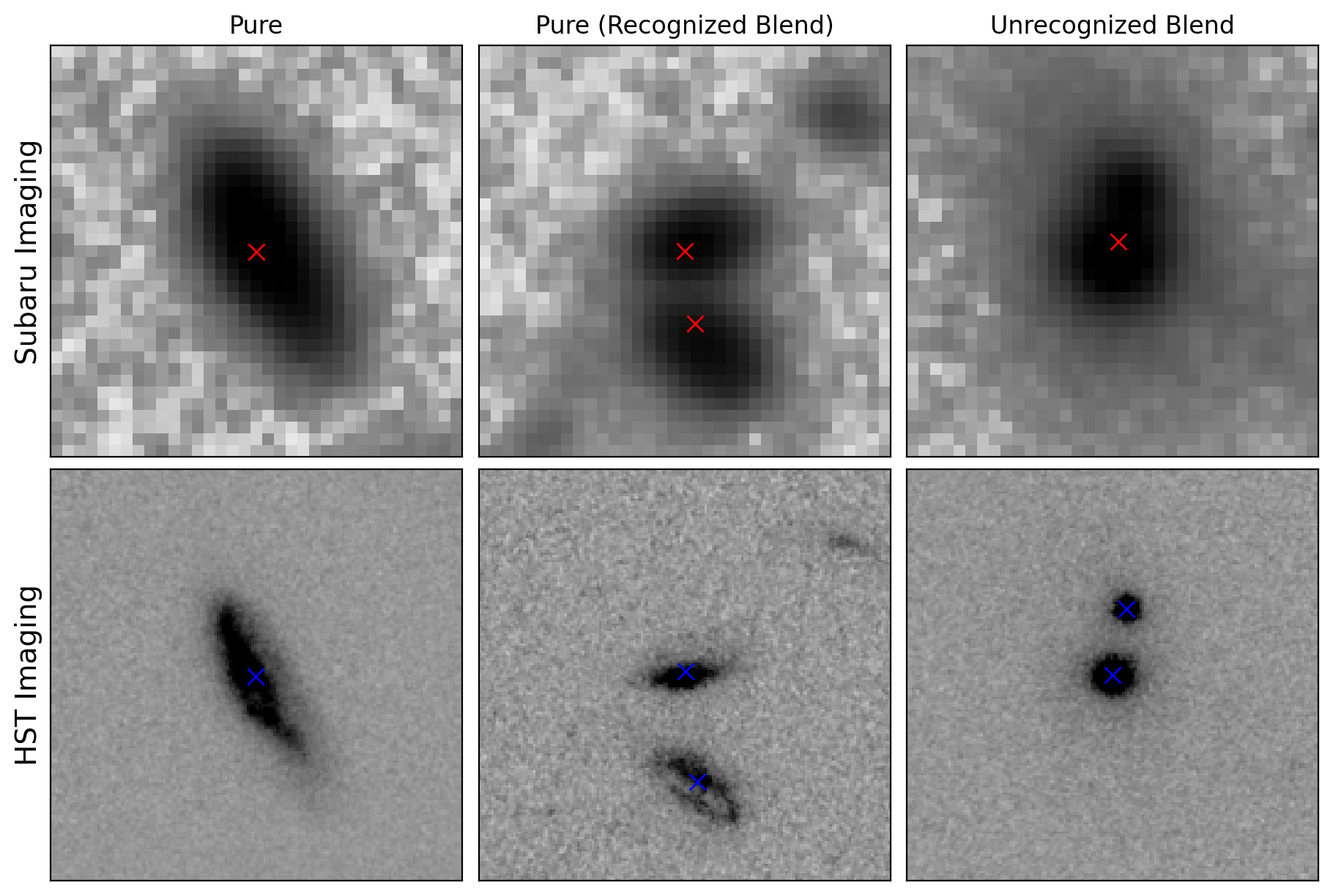}
    \caption{\edit{Example of various blend/non-blend configurations with Subaru $i$-band imaging on top and HST $F814W$ imaging below. From left to right we have an example of a pure galaxy, a pair of pure galaxies (recognized blend), and an unrecognized blend. The detection centers are overlaid as red crosses for Subaru and blue crosses for HST. }}
    \label{fig:hst_subaru_blendex}
\end{figure*}

Fortunately, the majority of blends can be recognized by source detection algorithms and deblended with dedicated tools such as \texttt{SCARLET} \citep{scarlet}. 
However, \citet{dawson} show that 14\% of detections could be ``unrecognized blends'' -- closely overlapping sources detected as one object \citep[an ``ambiguous blends'' in][]{dawson} -- for the LSST ``gold sample'' at $i\sim 25.3$. 
These blended sources are unrecognizable as blends based on ground-based observations, but can consistently be identified by high-resolution space-based instruments such as the Hubble Space Telescope (HST), the James Webb Space Telescope, \edit{and the Euclid Space Telescope.} 
\edit{Examples of these objects in ground-based (Subaru) and space-based (HST) imaging are shown in Fig.~\ref{fig:hst_subaru_blendex} showcasing an isolated galaxy, a deblended galaxy, and an unrecognized blend. }
More recently, \citet{roman_synth} find that, based on a synthetic image simulation of a 20 deg$^2$ overlapping region between LSST and the Nancy Grace Roman Space Telescope\footnote{\url{https://roman.gsfc.nasa.gov/}} High-Latitude Imaging Survey (HLIS), 20-30\% of $i<25$ objects identified by LSST are multiple objects in Roman images.

The impact of blending is a key challenge for deep ground-based surveys like LSST. 
Specifically, blending affects two essential measurements for weak lensing cosmology: shapes of galaxies (cosmic shear), and redshift estimate (photometric redshift, or photo-\textit{z}).
\citet{des_bl} find that blending-related effects are the dominant contribution to the mean multiplicative shear bias of approximately 2\,\% in DES.
\citet{rachel} emphasizes that the impact of blending on photometric redshift estimation is particularly difficult, since the spectroscopic redshift (spec-\textit{z}) selection for samples used for training and calibrating photo-\textit{z} may have different unrecognized blend rates than the general galaxy population. 
\citet{cunha} show that, without carefully matching the redshift distribution between the weak lensing sample and the calibration spectroscopic redshift sample, a spectroscopic sample with 2\% incorrect redshifts can cause a $\sim$15\,\% error in \textit{w} for DES (see Table 2 therein).
\edit{The DEEP2 spectroscopic survey \citep{Newman_DEEP2} estimated a 1 to 2\% contamination rate for galaxies with $z < 1$ suggesting that current spectroscopic training samples are compromised at a non-negligible level for both current and future surveys.}
More recently, based on an emulated mock catalog of 10\,254 square degrees and an equivalent depth of 1.3 years of LSST, \citet{erfan} observe a $\sim 0.025$ decrease in the derived structure growth parameter $S_8 = \sigma_8 (\Omega_{\mathrm{m}}/0.3)^{0.5}$ due to blending, corresponding to a $\sim2\sigma$ shift, where $\sigma$ is the statistical uncertainty on the shift.
This shift is comparable to the ``tension'' between state-of-the-art cosmic shear measurements \citep[e.g.,][]{des_y3_cs1, des_y3_cs2} and the \textit{Planck} Cosmic Microwave Background result \citep{planck_2018}.

\edit{Previous attempts at detecting unrecognized blends have typically focused on image level analysis and are validated on simulation data. This includes implementations like DeepDISC \citep{deepdisc_old}, using gaussian processes \citep{buchanan_gp}, or residual analysis after deblending \citep{kamath202thesis}. 
DeepDISC was recently ran on HSC data in \citet{deepdisc_new}, but focuses on the general detection of objects with limited attention to unrecognized blends.
We focus on the direct classification of unrecognized blends and the use of observational data for training, testing, and validation. 
}

When two galaxies are blended in projection, there is generally no correlation between their redshifts or host environments, and thus no correlation in their colors.
Therefore, the color of an unrecognized blend can be a combination of colors of any two (or more) types of galaxies, which is sometimes impossible to produce under normal baryonic physics. In addition, unrecognized blends might show unique morphology, or color/morphology combinations.
While we expect unrecognized blends to be larger, brighter, and have exotic colors on average due to partial overlap, we do not expect there to be a simple catalog cut to isolate them.
Therefore, we investigate the use of Machine Learning (ML) based on broad-band photometry from ground-based imaging to detect these features at the catalog level.
Specifically, we investigate the use of Self Organizing Maps \citep[SOM;][]{kohonen2, kind, dan_som}, Random Forest classifiers \citep[RF;][]{Breiman1996}, k-Nearest Neighbors \citep[k-NN;][]{ogknn_1, ogknn_2}, and several unsupervised anomaly detection methods for this study.

We test the methods on the COSMOS data set \citep{laigle} using two sets of features: 6-band photometry (\textit{uBVr$i^{\p}z^{\pp}$}) mimicing Rubin-LSST observations, or 9-band photometry (\textit{uBVr$i^{\p}z^{\pp}$YJH}) following the SOM configuration of C3R2 \citep{dan_som}.  We find that k-NN, RF, and SOM are all viable options to effectively identify unrecognized blends while many of the anomaly detection methods are comparable to a random selection.

In addition, we measure the impact on photo-z from removing unrecognized blends. Generally, we find that the removal of unrecognized blends reduces the rate of catastrophic outliers in photo-\textit{z}. Furthermore, we find that algorithms targeting color outliers are better at detecting photo-z outliers than algorithms targeting unrecognized blends.

This paper is organized as follows. In Sect.~\ref{sec:data}  we describe the data sets used in this work together with the catalog matching method. In Sect.~\ref{sec:method} we briefly introduce the Machine Learning algorithms and describe in detail our methodology for identifying unrecognized blends and catastrophic outliers in photo-\textit{z}. Our results are presented in Sect.~\ref{sec:res} and we discuss possible implications to weak lensing cosmology in Sect.~\ref{subsec:dis_app}. A summary of this work is provided in Sect.~\ref{sec:sum}.

%% file: 2.1.tex
\section{Data Description}
\label{sec:data}

We use the COSMOS data set \citep[Sect.~\ref{subsec:cos};][]{laigle} as the base catalog for this work. It is a combined catalog of data from several ground-based facilities. 
We match the COSMOS data to the HST catalog of the same region \citep[Sect.~\ref{subsec:hst};][]{hst1, hst2} to identify the unrecognized blends for the training and validation of our algorithms. We also match the COSMOS data to spectroscopic redshifts in the COSMOS region (Sect.~\ref{subsec:spec}) to study the impact of blends on photo-\textit{z}.
\edit{Our main catalog contains \textit{uBVr$i^{\p}z^{\pp}$YJH} bands along with the half-light radius ($i^{+}$  $\mathrm{FLUX\_RADIUS}$).}


We emphasize that the data selection on the ground-based COSMOS data set is only applied \textbf{after} the cross-matching between the COSMOS and HST catalogs.
This is important for preventing mis-identification of unrecognized blends and is discussed in more detail in the following section.

\subsection{The COSMOS Data Set}
\label{subsec:cos}
The COSMOS data set \citep{laigle} contains 30-band photometry and precise photometric redshifts based on the 30-band photometry for more than half a million objects over the 2 deg$^2$ COSMOS field at a $3\sigma$ depth of $i^{+} < 26.2$.
In this work, we utilize 9 band photometry data in COSMOS (\textit{u}-band from CFHT, \textit{BVr$i^{\p}z^{\pp}$}-bands from Subaru, and \textit{YJH}-bands from UltraVISTA) to mimic  LSST optical-band observations with Euclid synergies \citep{euclid_2011}. 
The COSMOS optical sources are detected on co-added images of $z^{++}YJHK_s$ bands; \texttt{SExtractor} \citep{sextracotr} is used for detection and deblending. 
The LSST pipeline will use more sophisticated tools to detect \citep{hsc_pipe} and deblend sources \citep[e.g., \texttt{SCARLET};][]{scarlet}. However, \texttt{SCARLET} requires detected source locations as input and does not natively identify single sources that are unrecognized blends.

We apply \textbf{no} preliminary selection on this ground-based COSMOS data set before cross-matching with HST.
This is important for preventing some of the ``recognized blends'' to be mistaken as unrecognized blends. 
The recognized blends are pairs of close-by but deblended objects in the ground-based catalog, matched to two (or more) HST objects as shown in Fig.~\ref{fig:demo}. If one of the two objects has been removed from the ground-based catalog for any of the reasons described below, the other object will be matched to two HST objects and will be misidentified as an unrecognized blend. Possible causes for the first object to be missing from the ground-based catalog include but are not limited to the following:
\definecolor{dgreen}{rgb}{0.,0.59215686,0.65490196}
\definecolor{amber}{rgb}{1,0.67058824,0.25098039}
\begin{figure}
\centering
\setlength\extrarowheight{3pt}
\begin{tabular}{ @{}c@{}|@{}c@{} }  \hline

    \begin{tabular}{cc}
        \multicolumn{2}{c}{ {\color{dgreen}{\huge{---}}} Detection}  \\
        \multicolumn{2}{c}{ {\color{amber}{\huge{---}}} Truth{\color{white}{.......}}}      \\ 
    \end{tabular} & 

    \begin{tabular}{c r}
        \multicolumn{2}{c}{Recognized Blends?} \\
         No                 &           Yes    
    \end{tabular} \\ \hline

    \begin{tabular}{c c}
        \multirow{2}{*}{Unrec-BL?}  &  \raisebox{4.0\height}{No}  \\
                                    &  \raisebox{-3.0\height}{Yes}  \\
    \end{tabular} &

    \begin{tabular}{c c}
        \includegraphics[scale=0.3]{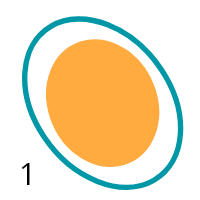} & \includegraphics[scale=0.3]{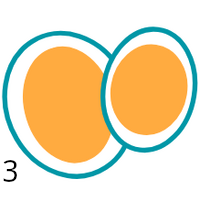}  \\
        \includegraphics[scale=0.3]{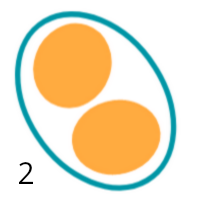} & \includegraphics[scale=0.3]{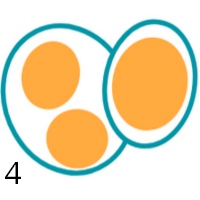}
    \end{tabular} \\ \hline
    
\end{tabular}
\caption{Illustration of different matching scenarios. Case 1: The ground detection matches a space detection with no other sources in the vicinity (1-1 match). This is a ``pure'' source. Case 2: The ground detection matches two space sources (1-2 match). This is an unrecognized blend. Case 3: Recognized blends, where two ground-detections are closeby but deblended, and matched to two space sources. They are also both pure sources, but can be mistaken as unrecognized blends if we only match between the two catalogs in one run, or if one of the ground-detections is missing due to data quality cuts. We use a two-step matching algorithm to avoid such situation (Sect.~\ref{subsec:match}), and only apply selections in the ground-based catalog  after matching. Case 4: One (or both) of the deblended component is itself an unrecognized blend. Our two-step matching method can classify this situation correctly.} 
\label{fig:demo}
\end{figure}

\begin{enumerate}[leftmargin=*,labelindent=10pt, label=\arabic*.]
    \item \textbf{Star/galaxy separation}: The first object is a star and the other a galaxy. If the star/galaxy selection is applied to the ground-based catalog before matching, the galaxy will match to both the star and the galaxy from HST. Notice that it is not feasible to apply star/galaxy separation to the HST catalog and match only between galaxies, since galaxies do blend with stars.
    \item \textbf{Magnitude/SNR cut}: The first object is rejected by the magnitude cut (e.g., $i^{\p}=24.6$ for a cut at $i^{\p}$ < 24.5) while the other survives (e.g., $i^{\p}$ = 24.4). Since we do consider situations where a galaxy blends with a source up to 2 magnitudes fainter, the brighter object will be mistaken as an unrecognized blend. 
    \item \textbf{Photometry quality flags}: One of the objects with bad photometry is excluded, and the other is mistaken as an unrecognized blends. This means that all sources need to be kept for matching as long as they are real detections. This includes quality cuts such as CCD saturation, bleeding, sources impacted by bad pixels/columns, CCD edge truncation, and any other instrumental effects.
\end{enumerate}

In addition to preserving the selection after matching, extra care is needed in order to identify unrecognized blends. We present the matching algorithm in Sect.~\ref{subsec:match}, and the selection after matching in Sect.~\ref{subsec:cuts_after_match}.

%% file: 2.2.tex
\subsection{The HST Data Set}
\label{subsec:hst}
We use the COSMOS HST/ACS data set \citep[hereafter the HST data set;][]{hst1, hst2} as the truth for blends. 
The HST data set was obtained with the Hubble Space Telescope in the F814W filter and corresponds to a field that is 1.64 deg$^2$ in the COSMOS area at a $5\sigma$ depth of 27.2. 

In contrast to the ground-based catalog, we can safely place cuts on the HST data set before matching. This is enabled by our two-step matching algorithm detailed in Sect.~\ref{subsec:match}. We apply selections on magnitude, size (for diffused sources), and a morphology-based flag for fake sources. See App.~\ref{sec:sel_hst} for more details. As noted earlier, we do not remove stars from the HST data set, since stars can blend with galaxies. For the same reason, we keep HST data objects down to $m_{814}$ < 26.5 to study blending with a fainter source up to 2 magnitudes fainter, corresponding to a flux contamination of at least 16\%, than the target depth of $i^{\p}<24.5$ (more details in Sect.~\ref{subsec:match}).
In total, 719\,935 sources passed the above cuts. 

In the matching procedure to identify unrecognized blends, we use the object size information from both the ground and the HST catalog (see Sect.~\ref{subsec:match}). Therefore, we calculate the effective size $r_H^\prime$ (as observed by Subaru) of all HST sources. It is achieved by convolving the HST flux radius $r_H$ with the ground-based PSF, approximated as
\begin{equation}
    \left(r_H^\prime\right)^2 = (0.049\times r_H)^2 + \Delta_{\mathrm{PSF}}^2,
    \label{eqn:eff_size}
\end{equation}
where
\begin{equation}
    \Delta_{\mathrm{PSF}}^2 = (3.328\times0.2)^2 - (2.28\times0.049)^2
\end{equation}
is the squared difference of the Subaru PSF and the HST PSF in arcseconds. Here the Subaru Prime Focus Camera (Suprime-Cam) pixel 
scale\footnote{\url{https://subarutelescope.org/en/about/instrument/suprime_cam/index.html}} is 0.2 arcsec/pix, and the HST ACS pixel scale\footnote{\url{https://hst-docs.stsci.edu/acsdhb/chapter-1-acs-overview/1-1-instrument-design-and-capabilities}} is 0.049 arcsec/pix. The PSF sizes (in pixels) for both instruments (3.328 for Subaru and 2.28 for HST) are estimated as the mode of the flux radius distribution of stars.

%% file: 2.3.tex
\subsection{Catalog Matching}
\label{subsec:match}
We match the ground-based COSMOS catalog to the HST data to identify blended sources.
A naive approach would be to count the number of matches for each ground source: those that match to only one HST source are not blended (``pure''), and those that match to multiple HST sources are blended.
However, this simple approach will likely misidentify recognized blends as unrecognized, since each of the recognized blends can be matched to two close-by space detections, as shown in Fig.~\ref{fig:demo}. In order to avoid such situations, we follow a similar approach as in \citet{dawson} with a two-step matching algorithm.

\begin{enumerate}[leftmargin=*,labelindent=10pt,label=\arabic*.]
    \item \textbf{Primary Match}: A primary match is the closest HST object within 2 arcsec of a COSMOS object.
    \item \textbf{Additional Match}: An additional match is any HST source that is non-primary, and satisfies \edit{both}:
    \begin{itemize}
        \item $\Theta / (r_C + r_H^\prime) < 1.5$. Here $\Theta$ is the angular separation between the COSMOS object and its additional match, $r_C$ is the flux radius of the COSMOS galaxy, and $r_H^\prime$ is the effective flux radius of the additional match convolved to the ground PSF (Eq.~\ref{eqn:eff_size}). 
        \item $m_{814} - i^{\p} \leq 2$. Here $m_{814}$ is the HST F814W filter magnitude, and $i^{\p}$ is the ground-based COSMOS magnitude. We allow galaxies to blend with fainter sources (including stars) with a magnitude difference less than 2 \edit{(at least a 16\% flux contamination)}. The magnitude difference is calculated between the ground-based object and its additional match from HST. In case there are multiple additional matches, the difference is calculated for the brightest additional matches to yield the smallest magnitude difference.
    \end{itemize}
\end{enumerate}
We define COSMOS galaxies as \edit{``\textbf{pure}''} (up to the HST-ACS resolution) when they have only a primary match. Those that have at least one additional match are marked as unrecognized blends. 
Note that 0.8\,\% of COSMOS galaxies share their primary match with another galaxy, \edit{which is an edge case not classified by Fig.~\ref{fig:demo} corresponding to a 2-1 match (possibly caused by shredding). But since none of them are unrecognized blends, we treat them as normal pure sources.}
\edit{\textbf{Unrecognized blends}} are further classified depending on the magnitude difference $\Delta_i = m_{814} - i^{\p}$, where sources with $1 \leq \Delta_i \leq 2$ are called \edit{``\underline{weak blends}''}, and those with $\Delta_i < 1$ are called \edit{``\underline{strong blends}''}. 
\edit{The} weak blends have at least $\sim$16\,\% of the flux contribution from their additional match(es), while the strong blends have at least $\sim$40\,\% flux contribution from the extra component(s). 
\edit{If $\Delta_i > 2$, i.e. the secondary component contributes less than 16\% of the total flux as estimated from the HST objects, we assume that the impact from the contaminant on photometry and shape measurements will be negligible.
Finally, we note that an object being labeled pure or an unrecognized blend is independent of it has been deblended or not, as shown in Fig.~\ref{fig:demo}. Definitions of these terms can be found in Sect.~\ref{subsec:sample}}

The combination of the two sets, $\Delta_i \leq 2$, is called ``\textbf{all blends}''.
From here on, ``unrecognized blends'' will refer to the ``all blends'' sample unless specified otherwise. Note that the pure sample is defined as the complementary set of the all blends sample; i.e., no additional match(es) within 2 magnitude difference.

%% file: 2.4.tex
\subsection{Selections After Matching}
\label{subsec:cuts_after_match}
After matching, we apply a series of cuts to the joint sample to select sources with robust photometry.
\edit{They are all selections on columns from the ground-based COSMOS data such as removing low surface brightness objects, saturated sources, and masked areas (more details in App.~\ref{sec:sel_ground}).}
In addition, we remove sources that are outside of the HST coverage, which results in a ``base'' COSMOS sample consisting of 138\,844 high S/N sources at a depth of $i^{\p}<24.5$.
\edit{These cuts mimic a ``Gold Sample'' like those found in cosmological surveys \citep{desY3_gold, desY6_gold, kids_gold} which better prepares us for application to LSST cosmology.
This limit, in conjunction with the HST completeness limit at 26.5 \citep{hst_cosmos_wl}, is further motivation for the $\Delta_i \leq 2$ cut-off written above.
}

\edit{We note that since our main aim here is weak lensing, we keep only objects classified as galaxies in the COSMOS catalog, but that the star-galaxy separation in COSMOS differs substantially from that in LSST.  The COSMOS catalog bases the star-galaxy separation on spectral fitting of 30 bands spanning from GALEX NUV to Spitzer IR imaging, whereas the LSST Science Pipeline star-galaxy separation is primarily morphology-based (and limited to optical bands).  In addition, deep Chandra imaging of the COSMOS field allows the identification of X-ray AGNs, which removes 1668 objects that otherwise pass our photometry cuts (171 X-ray AGN do not pass these cuts).}

Out of the 138\,844 base sample, 60\,390 sources are labeled by \texttt{SExtractor} as isolated in all bands (see item 2 above). We refer to these sources as the ``Non-deblended sample'', to differentiate them from the ``pure'' sample as defined in Sect.~\ref{subsec:match} and to reflect the fact that even when they are isolated by ground detection, they can still be unrecognized blends. The remaining 78\,454 sources are called the ``Deblended sample''. Notice that the Non-deblended sample is based on ground source detections only.
Note that we do not cut on flux errors, nor do we exclude non-detections in any bands\edit{, but we do remove all non-observations in at least one band (there are only 8 of them in total)}.
We adopt the Lupton {\it asinh} magnitude \citep{lupton}, or ``Luptitudes'', a magnitude system designed to work even at low or even negative fluxes (see App.~\ref{sec:appb} for how we calculated Luptitudes for the base COSMOS sample). 

After all of the above selections are applied, we find that more than 99\,\% of galaxies in the base COSMOS sample have a primary match with HST, and around 17\,\% of them have additional match(es), and thus are unrecognized blends (weak or strong). This is similar to 14\,\% unrecognized blends found in \citet{dawson} for the LSST gold sample at $i\sim 25.3$ and the 20\,\% to 30\,\% of unrecognized blends at $i<25$ for LSST from \citet{roman_synth}.
We will refer to this cross-matched COSMOS$\times$HST sample as the ``labelled sample'', meaning the unrecognized blends are labelled for training and testing.

%% file: 2.5.tex
\subsection{Spectroscopy Data Sets}
\label{subsec:spec}
We assemble the spectroscopic redshift catalog from a range of data sets covering the COSMOS field, including zCOSMOS \citep{zcosmos1, zcosmos2}, DEIMOS \citep{deimos1, deimos2}, C3R2 \citep{dan1, dan2, stanford, lbt}, VUDS \citep{vuds1, vuds2}, and DESI EDR \citep{desi_edr}. Detailed selections on these data sets can be found in APPENDIX~\ref{sec:sel_spec}.

We combine all catalogs into one, and check for duplicated sources by matching their coordinates within 1 arcsec. We find 5\,705 duplicated sources, out of which 5\,525 sources are considered to be the same sources at the requirement that their redshifts agree within $z<0.01$ among different catalogs. The remaining 180 sources with larger than 0.01 redshift difference are discarded. Finally, we obtain a spectroscopic redshift catalog of 27\,900 sources within the COSMOS field, which we refer to as the ``spec-\textit{z} sample''.  We then match the labelled COSMOS sample to the spec-\textit{z} sample within 1 arcsec. In total, 20\,154 spectra out of  27\,900 are matched to the COSMOS sample, \edit{ where most of the missing spectra are matched to the excluded sources under the selections for the labelled COSMOS sample (App.~\ref{sec:sel_ground}) including 894 AGNs, and} a few hundred unmatched spectra falling in the outskirts or masked regions of the COSMOS footprint.

%% file: 2.6.tex
\subsection{Sample Statistics}
\label{subsec:sample}
Some summary statistics of the labelled COSMOS sample are listed in Table~\ref{tab:match}. 
\begin{table*}
	\caption{Summary of the labelled COSMOS sample and all subsamples. The Deblended/Non-Deblended separation of the Total sample is based on the ground-based detection, and the Pure/BL separation is based on matching with HST. The subsamples are defined in Sect.~\ref{subsec:match}. The photo-\textit{z} outliers in this table are defined between the COSMOS 30-band photo-\textit{z} ($z_{\mathrm{30}}$) and spec-\textit{z} ($z_{\mathrm{spec}}$) by $|z_{\mathrm{30}} - z_{\mathrm{spec}}| >  0.15(1+z_{\mathrm{spec}})$, and the outlier fraction is the fraction of photo-\textit{z} outliers over the spec-match subsample. While AGNs are removed from the actual sample, we include their counts .}
	\label{tab:match}
	\centering
	\begin{tabular}{rccccccc} \toprule
                              &  Total   &  Deblended   & Non-Deblended &  Pure   &   All BL  & Strong BL & AGN  \\
		Num. Sources          &  138844  &     78454 (56.5\%)    &     60390 (43.5\%)     & 115338 (83.1\%)  &   22233 (16.0\%)   &   5326 (3.8\%) & 1668    \\
		Pure Fraction         &  83.1\% &     83.3\%  &     82.8\%   &  100\%  &    0\%    &    0\%  & 86.5\%   \\
		Num. Spectra Match    &  20154   &     14054    &     6100      &  18370  &    1708   &   419  & 894    \\
		Num. Photo-z Outliers &   628    &      504     &      124      &   484   &    139    &   48   & 95    \\
		Outlier Fraction      &  3.1\%  &     3.6\%   &     2.0\%    &  2.6\% &   8.1\%  &   11.5\%  & 10.6\%  \\
        \bottomrule
    \end{tabular}
\end{table*}
\edit{Our definitions for these columns are:}
\begin{enumerate}[leftmargin=*,labelindent=3pt,label=\arabic*.]
    \item[] \textbf{Deblended}: Sources marked as blended in any of the \textit{uBVr$i^{\p}z^{\pp}$YJH} bands of COSMOS, but with fluxes successfully recovered. They can be either recognized blends (case 3 in Fig.~\ref{fig:demo}) or unrecognized blends (case 4).
    \item[] \textbf{Non-Deblended}: Sources isolated (therefore not deblended) in all of the \textit{uBVr$i^{\p}z^{\pp}$YJH} bands. They are the ``purest'' sample definable with ground-based observations, but they can be unrecognized blends (case 1 or 2).
    \item[] \textbf{Pure}: Sources with only a primary match to the HST catalog. They are isolated or deblended clean sources up to the HST resolution (case 1 or 3).
    \item[] \textbf{Unrecognized Blends (BL)}: Sources with a primary match and at least one additional match. They are certainly blended sources, but not recognized as a blend by the ground-based detection (case 2 or 4). They are further divided into \textbf{weak} or strong \textbf{blends} depending on the flux contribution from additional matches.
\end{enumerate}
Notice that the labelled sample is divided roughly equally into the deblended and non-deblended categories. Both categories consist of 
about 83\,\% pure sources and 17\,\% unrecognized blends. The numbers likely depend on the details of the ground-based detection/deblending configuration of COSMOS \citep[low threshold, high contiguous number of pixels with {\tt SExtractor}; see Sect.~3.1.1 of][]{laigle}, but in this case, it does not capture information about unrecognized blends. In this work, however, we will show that the situation can be improved by adding the color (and size) information. 

We summarize quantities associated with ``photo-\textit{z} outliers'' in Table~\ref{tab:match}, which are galaxies with a large disagreement between the COSMOS 30-band photo-\textit{z} ($z_{\textrm{30}}$) \citep{laigle} and spec-\textit{z} ($z_{\textrm{spec}}$). Specifically, we define photo-\textit{z} outliers if 
\begin{equation}\label{eq:spec_z}
 |z_{\mathrm{30}} - z_{\mathrm{spec}}| >  0.15(1+z_{\mathrm{spec}}).   
\end{equation}
Notice that the deblended sample has a higher photo-\textit{z} outlier rate than the non-deblended sample, which provides clear evidence that photo-\textit{z} is impacted by imperfect deblending.  The pure sample consists of both the deblended sources and the non-deblended galaxies, thus its outlier rate is in between the two. The unrecognized blends samples have a much higher outlier rate, reflecting their unphysical colors due to the mixture of components. 
\edit{}
Later in this work, we estimate SOM-based photometric redshifts based on 6 or 9 bands and size (Sect.~\ref{subsec:somz}), and study the identification of the 6-band and 9-band photo-\textit{z} outliers.

\begin{figure*}
	\centering
        \includegraphics[width=.9\columnwidth]{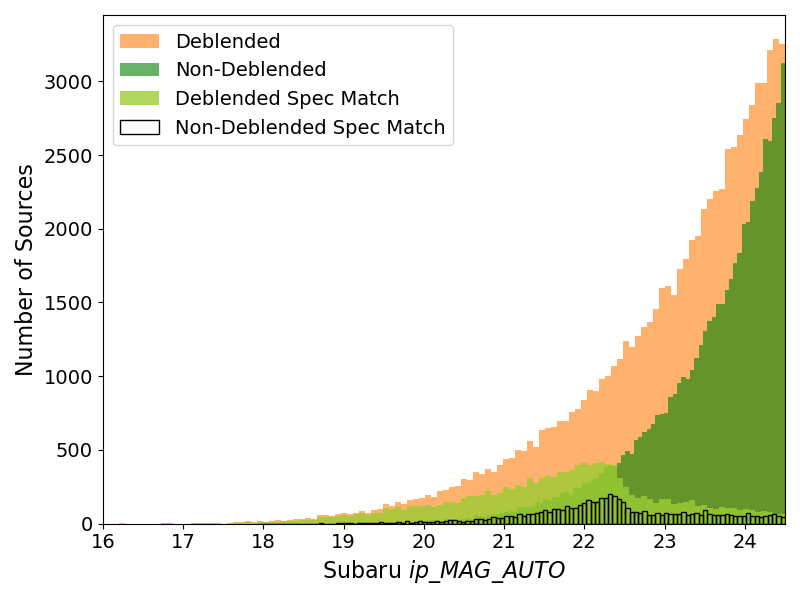}
        \includegraphics[width=.9\columnwidth]{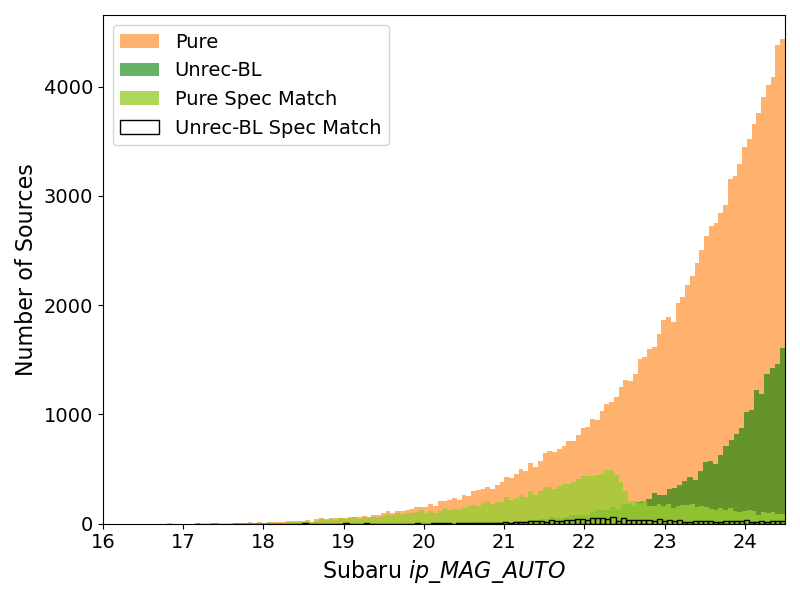}
\caption{Magnitude distribution of the labelled COSMOS sample. Left: dividing the sample into Deblended and Non-Deblended sub-samples by ground-based detection. Right: dividing the sample into pure and unrecognized blends by matching with HST. The magnitude distributions of sub-samples with a spectrum match are shown in both panels. All samples are defined in Sect.~\ref{subsec:cuts_after_match} and Sect.~\ref{subsec:spec}. Plotted magnitude is the derived Luptitudes from Subaru flux measurements. The spectrum-matched sources with $i^{\p} > 22.5$ are mostly contribution from the C3R2 catalog. The unrecognized blends in the right panel are weak blends (see Sect.~\ref{subsec:match}). Summary statistics of these samples are shown in Table~\ref{tab:match}.} 
	\label{fig:mag_hist}
\end{figure*}

Shown in Fig.~\ref{fig:mag_hist} are the magnitude distributions of the labelled COSMOS samples as well as their spectrum matches. A few features are observed. First, in the left panel, notice that the deblended sample contains more bright sources than the non-deblended sample. This is because brighter sources tend to be bigger and are more likely to blend with neighbors. For the same reason, the deblended sample has more spectrum matches than the non-deblended sample because the spectroscopic surveys select brighter sources. Also notice that the spectrum-matched samples peak at $i^{\p} \sim 22.5$ with a long tail due to our spectrum library being assembled from several surveys. The sources fainter than 22.5 mainly come from C3R2 and DESI EDR.

%% file: 3.0.tex
\section{Methods}
\label{sec:method}
Our key assumption is that some of the unrecognized blends have different colors from the isolated sources and that this feature can be identified with Machine Learning.
\edit{In total we use 10 features to inform the models: $u-B$, $B-V$, $V-r$, $r-i^+$,  $i^+ - z^{++}$, $z^{++} - Y$, $Y-J$, $J-H$, $i^+$, and flux radius.}
We focus on algorithms that have found success in astronomy such as self-organizing maps (SOM), random forests (RF), k-Nearest Neighbors (k-NN), and several anomaly detection methods.
\edit{The subsequent subsections will introduce each of these methods along with specific implementation details that are relevant for our unrecognized blend classification.
Details on anomaly detection methods are split between Sect.~\ref{subsec:anomaly} and App.~\ref{sec:anom}.}

%% file: 3.1.tex
\subsection{The SOM Configurations}
\label{subsec:conf}
\edit{A SOM is an unsupervised neural network that can reduce the dimensionality of a dataset by mapping to a fixed number of cells on a grid of an arbitrary shape: rectangle, hexagon, triangle, etc.}
We define our map on a unit sphere tiled with equal-area cells \citep[healpix\footnote{\url{http://healpix.sourceforge.net}};][]{hp1, hp2} to avoid boundary effects (special edge or corner cells). 
Each cell $k$ is associated with a weight vector $\vec{\omega}_k$, which has the same dimension as the features in the data.
In this study, we primarily use all 8 colors of the 9 bands plus 1 magnitude ($i^{\p}$) from the labelled COSMOS sample, together with 1 size measurement (flux radius).
The data features are re-scaled to be centered at 0 with a standard deviation of 1, and the components of the weight vectors are randomly initialized in $[-1,+1]$.

\edit{The training process takes one pure object, drawn randomly with replacement, from the training sample (defined later) to train the SOM.}
The randomly selected object is mapped onto the SOM cell whose weight is ``closest'' to its features. Euclidean distance is often used as the metric (see Sect.~\ref{subsec:alt_conf}); however, in this study, we use $\chi^2$ distance to incorporate the uncertainty in photometry:
\begin{equation}
    \chi_k^2 = \sum_i \frac{(F_i - \omega_{k,i})^2}{\sigma_i^2 + 1},
    \label{eqn:dist}
\end{equation}
where $\vec{F}$ is the feature vector of the source (8 colors, 1 magnitude and 1 size) whose components are indexed by $i$, and $\sigma_i$ is the uncertainty in the features. For colors and magnitude, the uncertainty is propagated from the magnitudes and re-scaled by the same amount as the features. A constant re-scaled uncertainty of 0.2 is set for the size measurement. The $\chi^2$ distance is calculated with both the feature uncertainty $\sigma_i$ and a population-level variance of 1. In some rare cases where one or two features are missing from a source (non-observation in a band), the distance is calculated with only the available features, and boosted proportionally to account for the missing features. The cell with minimum distance $\chi^2_k$ is often referred to as the ``best-match cell'' for the source.

Once the galaxy is placed on its best-match cell, all SOM cells shift their weights towards the feature vector of the galaxy:
\begin{equation}
    \vec{\omega}_k(t+1) = \vec{\omega}_k(t) + \alpha(t) H_{b,k}(t)\left[ \vec{F}(t) - \vec{\omega_k}(t) \right],
\end{equation}
where $\vec{\omega}_k(t)$ is the weight of cell $k$ at step $t$ and $\vec{F}(t)$ is the feature of the source galaxy drawn for this step. The training process is controlled by two factors: the neighborhood function $H_{b,k}(t)$ and the learning-rate $\alpha(t)$.

The neighborhood function is designed to give larger shifts to neighboring cells than farther cells. It is defined with
\begin{equation}
    H_{b,k}(t) = \exp{\left[ -D^2_{b,k}/\sigma^2(t) \right]},
\end{equation}
where $D_{b,k}$ is often the Euclidean distance between cell $k$ and the best-match cell $b$ in pixel units, but in our map geometry, it is the angular distance between the cells in radians. The neighborhood function contains a width parameter $\sigma(t)$ for a decaying behavior over iterations, such that early-iteration galaxies affect farther cells on the map than later ones. It is defined as
\begin{equation}
    \sigma(t) = \sigma_0 \left( \frac{\sigma_f}{\sigma_0} \right)^{t/N_T},
\end{equation}
with an initial value $\sigma_0$ equal to the size of the entire map and an ending value $\sigma_f$ comparable to the size of a single cell. Here $N_T$ is the number of training iterations, which usually equals the sample size of training data.
However, we find that this number of iterations often results in a few empty cells (cells that are not best-match cell for any galaxies), which causes a problem for SOM\textit{z} (see Sect.~\ref{subsec:somz}). Therefore, we adopt an iteration number of 5 times the training sample size to guarantee color space coverage, which most often yields no empty cells. 
The learning-rate factor $\alpha(t)$ sets a global decaying rate for training in order for the map to converge eventually. It is defined with a starting value $\alpha_s$ and an ending value $\alpha_e$:
\begin{equation}
    \alpha(t) = \alpha_s \left( \frac{\alpha_e}{\alpha_s} \right)^{t/N_T}.
\end{equation}

We adopt a healpix resolution index 4, which has 3072 cells in total where each cell is approximately $8\times8$ deg$^2$. 
This results in an order of 20 galaxies per cell on average for the training sample.
Our SOM parameters are listed in Table~\ref{tab:para}. 

\begin{table}
	\caption{Parameter values of the SOM/RF configurations used in this work. For all variables see text for additional details. SOM: The $\alpha$ parameters are dimensionless scale factors, and the $\sigma$ parameters are in radians. Random Forest: N is number of trees in forest, $\textrm{N}_{\textrm{size}}$ refers to terminal node size, and $\textrm{N}_{\textrm{try}}$ is the number of variables tested at each split. k-NN: metric is the metric used to calculate distance between points, k is the number of neighbors, and weighting is the scheme used to weigh the labels of the k neighbors.}
	\label{tab:para}
	\centering
	\begin{tabular}{c c c c c c}
		\toprule
        \multicolumn{6}{c}{SOM Parameters}                                            \\
        \cmidrule(lr){1-6}
		Resolution & Cell Number & $\alpha_s$ & $\alpha_e$ & $\sigma_0$ & $\sigma_f$  \\
	          4      &     3072     &      0.9  &     0.5    &   $\pi$    &   0.052     \\ \midrule
           
        \multicolumn{3}{c}{RF Parameters}     &  \multicolumn{3}{c}{k-NN Parameters}  \\
        \cmidrule(lr){1-3}
        \cmidrule(lr){4-6}
          N        &  N$_{\textrm{size}}$   & \multicolumn{1}{c}{$\textrm{N}_{\textrm{try}}$}    &    Metric   &  k   & Weighting     \\
	       500      &           5           &             \multicolumn{1}{c}{3}                   &   Euclidean &  10  &  Uniform      \\ 
        \bottomrule \\
    \end{tabular}
\end{table} 

We use SOM to detect unrecognized blends and photo-z outliers in two approaches: the ``\textbf{blending cell approach}'', and the ``\textbf{distance cut approach}''. In both approaches, we split the labeled COSMOS sample into training and validation sets for cross-validation. The blending cell approach requires an additional split of an identification sample. They are defined as:
\begin{enumerate}[leftmargin=*,labelindent=5pt,label=\arabic*.]
    \item[] \textbf{The training sample}: A random half of the pure sample (one-to-one matches; see Sect.~\ref{subsec:match}), but not including sources with a spectroscopic match. This is because we have a limited number of spec-\textit{z}, which we reserve for validation purposes. To avoid unwanted selection effects, however, we do not simply exclude the spectroscopic matches from the training sample. Instead, we swap the sources with galaxies in the other half of the pure sample with similar properties and without a spectroscopic match. Specifically, we swap with a source with $\sum_i |\Delta b_i| < 10$, where $b_i \in$ \{\textit{u, B, V, r, $i^{\p}$, $z^{\pp}$, Y, J, H, $z_\mathrm{C30}$}\}. On average, this means that the source has less than 1 magnitude difference in each band and in the 30-band photometric redshift.
    \item[] \textbf{The identification sample}: A random half of the unrecognized blends not including sources with a spec-\textit{z} match. The same source-swap approach is used with the training sample with the swapped source coming from the other half unrecognized blends.
    \item[] \textbf{The validation sample}: The other half of the pure sample and the other half of the unrecognized blends. This sample contains all the spectroscopic matches and has the same proportion of pure and blended sources as the entire sample -- 17\,\% of unrecognized blends.
\end{enumerate}

\subsubsection{Blending cells}
\begin{figure*}
	\centering
        \includegraphics[width=0.98\columnwidth]{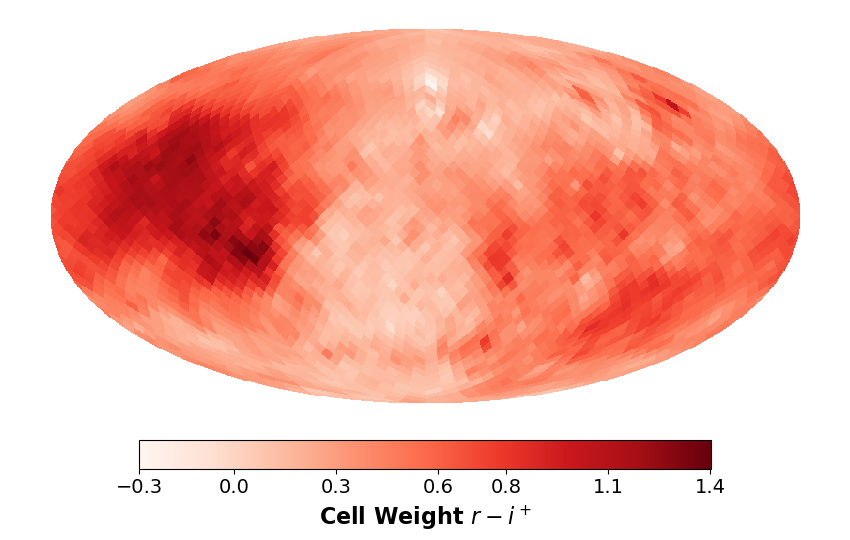}
        \includegraphics[width=0.98\columnwidth]{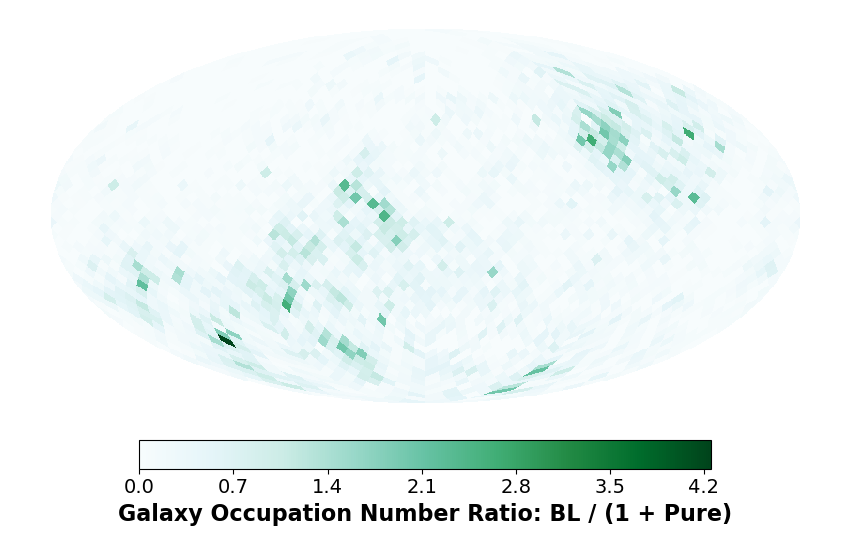}
	\caption{One example of the SOMs used in this work. The SOM is trained with the training sample described in Sect.~\ref{subsec:conf}. After training, both the identification sample and the training sample are mapped/re-mapped onto the SOM to identify blending cells. Left: The $r-i^+$ color of the 10-d SOM weight vectors. The weight vectors are scaled back to match the original colors of the training sample for illustration. Right: The ratio of occupation number of unrecognized blends (from the identification sample) to pure galaxies (from the re-mapped training sample) in each cell. The galaxy occupation number of a cell is simply the number of sources mapped onto the cell. The pure galaxy occupation is increased by 1 to avoid zero denumerators. We define the cells with high ratio of blends to pure sources as blending cells. Notice the apparent correlation between the two panels, which supports the assumption that colors contain information about unrecognized blends.} 
	\label{fig:som}
\end{figure*}

In the blending cell approach, the SOM is used as a classifier, where each cell is an individual class that spans a small sub-space in the 10-dimensional features.
The blending cell approach consists of three steps:
\begin{enumerate}[leftmargin=*,labelindent=10pt,label=\arabic*.]
\item Train a SOM with the training sample.
\item Map the identification sample onto the SOM and re-map the training sample onto the SOM. 
Count the occupation numbers in their best-match cells. Identify ``blending cells'' with a high ratio of blends to pure objects occupation numbers.
\edit{Note that we are not re-training the SOM, simply identifying cells with high blend occupancy.}
\item Map the validation sample onto the SOM. Identify the suspected unrecognized blends as sources mapped onto blending cells.
\end{enumerate}
To avoid statistical fluctuation in the training, we create 5 realizations of SOMs, where each realization is trained with a different set of randomly initialized cell weights as well as a different random split of the training/identification/validation samples.
A trained SOM is shown in Fig.~\ref{fig:som} as an example. The left panel shows the $r-i^+$ component of the SOM weight vectors, scaled back to match the original color of the sources. The right panel shows the occupation ratio of unrecognized blends from the identification sample to pure galaxies from the re-mapped training sample.
The pure galaxy occupation is increased by 1 to avoid zero denominator in the ratio. 
Notice that there is correlation between the occupation ratio and the SOM weight vector, evidence that the color information yields \edit{some} indication of unrecognized blends.

\subsubsection{Distance Cut}
\begin{figure}
	\centering
        \includegraphics[width=\columnwidth]{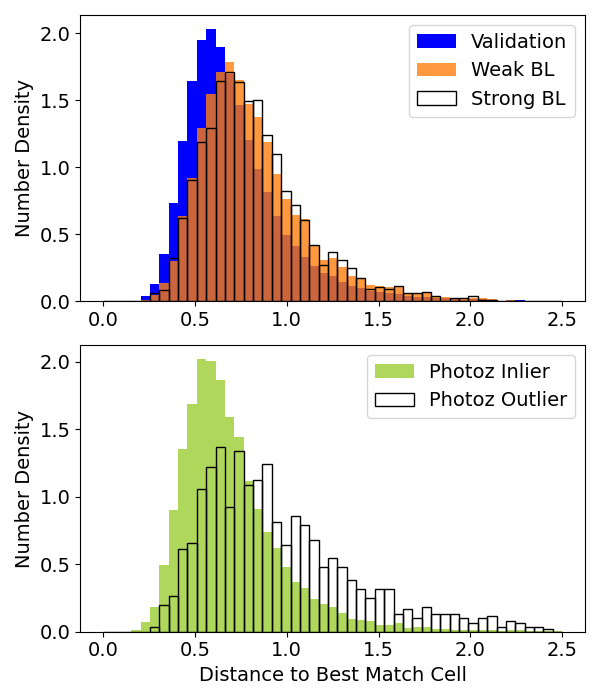}
	\caption{Distribution of $\chi^2$ distances between galaxies in the validation sample and their best-match SOM cells. The distances are calculated with the 10 scaled features (8 colors, $i^+$ band magnitude and object size) of galaxies using equation (\ref{eqn:dist}). The unrecognized blends populations tend to be moderately further away from their best-match cells. However, the photo-z outlier sample is significantly different from the inlier sample.} 
	\label{fig:dist_dist}
\end{figure}

The distance cut approach is more straight-forward: we use the $\chi^2$ distance of an object's features to the weight vector of its best-match cell (Eq.~\ref{eqn:dist}) as a blend indicator. We train the SOM on pure galaxies; therefore the cell weights reflect galaxy colors produced from baryonic physics. The blended objects, which often have unphysical colors, are then expected to have a larger $\chi^2$ distance to their best-match cells.  This distance distribution is shown in the upper panel of Fig.~\ref{fig:dist_dist}, where the unrecognized blends indeed have slightly larger distances overall.

The lower panel of Fig.~\ref{fig:dist_dist} shows the $\chi^2$ distance of the photo-\textit{z} inliers versus the ``outliers'', defined in Eq.~\ref{eq:spec_z}.
Here we show the 9-band SOM\textit{z} (calculated in Sect.~\ref{subsec:somz}) instead of the COSMOS 30-band photo-\textit{z}. The distance distribution is significantly different between the SOM\textit{z} outliers and the inliers. As a result, we expect the distance cut method to perform better in finding photo-\textit{z} outliers than in finding blends (Sect.~\ref{subsec:pz_res}). We then seek to remove objects according to their $\chi^2$ distances, from high to low, from the validation sample, and re-quantify the blends.
In order to directly compare with the blending cell approach, we use the same 5 SOM realizations as well as the same training/validation sets as used in that aspect. The identification sample is not used in the distance cut approach.

%% file: 3.2.tex
\subsection{Random Forest Algorithm}
\label{subsec:RF_config}

\begin{figure}
    \centering
    \includegraphics[width=1\linewidth]{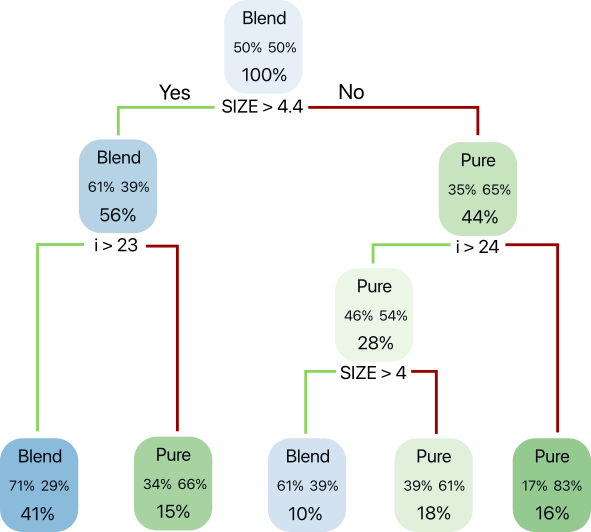}
    \caption{Example decision tree using a 50\% blend 50\% pure galaxy dataset. Each node consists of a label (top), either ``Blend'' or ``Pure'', the fraction of blends in the node (middle left), the fraction of pure galaxies in the node (middle right), and the fraction of the total sample in the node (bottom). The bottom row of nodes are the terminal nodes while the rows above can be split based on the decision underneath the node to further nodes in the tree.}
    \label{fig:dtree-ex}
\end{figure}

A RF is a supervised machine learning algorithm that consists of a collection of decision trees.
A decision tree is a hierarchical model that partitions dataspace through a series of decisions \citep{decision_tree1, decision_tree2}.
Each partition creates a node that is assigned a label based on the input training data and can be subdivided again to provide a more accurate estimate of the label.
The starting node, or root node, will be assigned a label that reflects the average label for the training data and subsequent splits will partition a node to create more consistent subsamples.
Each split, or decision, will be created in order to optimize some metric -- decreasing mean-square error or increasing Gini purity are examples for regression trees and classification trees, respectively.
If all training samples in a node share the same label, or any split would cause subsequent nodes to have fewer than $N_{\textrm{size}}$ samples in a subsequent node, the split is terminated and the node becomes a terminal node.
After a decision tree is trained, any test point simply follows the chain of decisions until it is put in a terminal node and is assigned the corresponding label. 
An example decision tree is shown in Fig.~\ref{fig:dtree-ex}

In order to ensure that the $N$ decision trees in a RF are distinct, some key features are implemented to decorrelate the trees, which usually improves collective predictive power.
Each tree is trained on a bootstrapped dataset that provides, on average, only $\sim60\;\%$ of the total dataset to a tree.
Furthermore, any split for a node in a tree will only test $N_\textrm{try}$ randomly selected features and, as stated above, split based on the feature that optimizes some user-defined metric.
These features can create individually poor trees that are collectively quite accurate \citep{Breiman1996}.
The hyperparameters used for the RF implementation can be found in Table~\ref{tab:para}.

In traditional classification problems with $N_l$ different labels, the final label is determined via plurality -- the winning label must have greater than $N/N_l$ votes, or simply the label with the most votes wins.
This scheme presents issues with unbalanced datasets where basic implementation of machine learning methods will often fall in a local minima of assigning every test point the popular, larger, label simply due to the training prior. 
Mitigating this effect is a rich field of study with a variety of solutions and we will implement threshold moving where the number of votes required to assign a certain label can be tuned.
We can set the threshold such that the less popular, smaller, label needs fewer votes and the larger label requires more votes.

For binary classification,  let $N_b$ be the number of trees that assign a datapoint the ``blend'' label.
We can vary the threshold and label accordingly:
\begin{align*}
        \frac{N_b}{N} < \mathtt{threshold} & \Rightarrow \textrm{Pure}  \\
        \frac{N_b}{N} \geq \mathtt{threshold} & \Rightarrow \textrm{Unrecognized blend}.
\end{align*}
This gives a similar flexibility as SOM where the final sample purity can be improved at the cost of a smaller final sample size (due to misidentifying pure galaxies).
Each datapoint is given a \verb|score|$_{\textrm{RF}}$ $ = \frac{N_b}{N}$ where \verb|score|$_{\textrm{RF}}\in(0,1)$ (0 corresponds to a ``pure'' object and 1 to a ``blended'' object) and the \verb|threshold| varies in the same range.


Our final model uses $N=500$ classification trees that are trained on colors, magnitudes, and flux radius with labels differentiating pure galaxies, weak unrecognized blends, and strong unrecognized blends.
While there are 3 labels, we primarily reduce this to binary classification by counting the number of trees voting for pure, $N_p$, and setting the number of votes for blends $N_b = N - N_p$.
We use this setup as it gives the best flexibility in our initial studies and, as we will discuss in Sect.~\ref{subsec:alt_conf}, there is a negligible change in performance if we used all three labels. 

Each tree in RF trains on a bootstrapped sample with the datapoints included for a tree denoted as ``in-bag'' while those excluded as ``out-of-bag'' (oob). 
The oob points can be processed after training on the relevant trees and act as a test set, in that they were not in the training set and have a defined label, bypassing the need for a traditional training-testing split. 
As stated above, bootstrapping recreates $\sim60\,\%$ of the initial sample, which means that $\sim40\,\%$ of trees will consider a datum oob.
However, for cross-validation with other methods we do use a well defined training testing split.
Specifically, for RF we use: 
\begin{enumerate}[leftmargin=*,labelindent=5pt,label=\arabic*.]
    \item[] \textbf{The training sample}: A random half of the pure sample and a random half of the unrecognized blends (see Sect.~\ref{subsec:match}), except that sources with a spectroscopic match are excluded following the same source swap algorithm as the SOM training sample (see Sect.~\ref{subsec:conf}). The RF training sample is a combination of the SOM training and SOM identification sample. 
    \item []\textbf{The validation sample}: The same validation sample as used in SOM. It contains the other half of the pure and unrecognized blends sample and has all the spectroscopic matches.
\end{enumerate}


This clear split between training and testing is a more accurate emulation of any future applications for observatories such as Rubin Observatory as it will have a well defined labeled training set and a much larger unlabeled set on incoming data. 
We have verified that there is similar performance between using the oob score and testing dataset score. 

%% file: 3.3.tex
\subsection{k-Nearest Neighbors}
\label{subsec:knn}

k-Nearest Neighbors (k-NN) is a non-parametric supervised algorithm, first described in \cite{ogknn_1}, that assigns a test datum a label based on the nearest $k$ neighbors in the training set. ``Nearest'' is typically measured using a Euclidean distance in the feature space; however there is some flexibility in the distance metric that allows the user to decide what is the best distance for a given dataset.
Additionally, a weighting scheme can be implemented that can weight the neighbors; common schemes include completely uniform or distance based scaling.
To account for any mismatch in scale of features -- and thus place increased weight on certain features -- the features are traditionally rescaled by subtracting the mean and dividing by the standard deviation. 

Similar to RF, we assign each datum a \verb|score|$_{\textrm{RF}}$ and use threshold moving to deal with imbalanced data.
The scores are set in a similar way: if a datum has $k_p$ votes for pure from its $k$ nearest neighbors, the score is 
\begin{equation*}
    \verb|score|_{\text{kNN}} = \frac{k - k_p}{k}.
\end{equation*}

For k-NN, we use the same training and validation sample as for RF. The training sample is used to define the pool of neighbors for any test datum in the validation sample. 
We have verified that the results are similar over all 5 realizations but for clarity only present the results from one below.
We have found consistent results for any $k > 5$ with marginal improvement on larger $k$ and the choice of metric providing little to no affect.

%% file: 3.4.tex
\subsection{Anomaly Detection}
\label{subsec:anomaly}
Anomaly detection generally refers to unsupervised learning algorithms that identify outliers in a dataset.
Given that we expect some blended objects to have irregular colors, we hope to find them as outliers in comparison to pure galaxies. 
The methods we have included are Elliptical Envelope (EE), Local Outlier Factor (LOF), Isolation Forest (iForest), and One-Class Support Vector Machine (One-Class SVM).
A general review of these techniques and many other methods can be found in \cite{anomaly1}, \cite{anomaly2}, and \cite{anomaly3}, and for astronomy specific uses, in \cite{astro_anomaly1}.
Further details on the anomaly detection methods and their performance can be found in App. ~\ref{sec:anom}.

In our main results we restrict ourselves to LOF as it performed the best among all the anomaly detection methods. 
LOF is an unsupervised algorithm that operates on the principle that outliers exist in local underdensities of data space \citep{breunig2000lof}. 
It is trained using SOM's training sample, consisting of pure galaxies only, and tested using the validation sample, made of a mix of pure galaxies and blended objects (see Sect.~\ref{subsec:conf} for definitions of samples).
The LOF algorithm produces a \verb|score|$_{\textrm{LOF}}$ which is a positive real number that is less than 1 for overdensities and greater than 1 for underdensities.
We rescale this to a range from 0 to 1 for each object in the test sample and then implement the same threshold moving scheme from RF to label blends.

%% file: 3.5.tex
\subsection{SOM\textit{z} And Outliers}
\label{subsec:somz}
To simulate a joint LSST plus Near Infrared data set or LSST alone, we measure photo-\textit{z} with a SOM (often referred to as SOM\textit{z}) with the 6 optical bands from COSMOS (\textit{uBVr$i^{\p}z^{\pp}$}) and with/without the 3 infrared bands (\textit{YJH}).
Our SOM\textit{z} is calculated with three steps:
\begin{enumerate}[leftmargin=*,labelindent=10pt,label=\arabic*.]
\item Train a SOM with the training sample.
\item Re-map the training sample onto the SOM. Calculate the mean of COSMOS 30-band photo-\textit{z} from training sample in each cell as SOM\textit{z} of the cell.
\item Map the validation sample onto the SOM. Assign each cell-SOM\textit{z} to validation objects in the cell.
\end{enumerate}

We use the same SOMs and data sets as used for blends detection (Sect.~\ref{subsec:conf}). 
The COSMOS 30-band photo-\textit{z} measurements are used as truth for the SOM\textit{z} training \edit{(pure galaxies only)} instead of our spectra sample in order to guarantee coverage in each SOM cell.
Table~\ref{tab:somz} lists relevant statistics for the validation sample.
We study two configurations for SOM\textit{z}: 10 features (optical+infrared+size) and 7 features (optical+size), which we refer to as 10-SOM\textit{z} and 7-SOM\textit{z}, respectively.
The definition of a photo-\textit{z} outlier in Eq.~\ref{eq:spec_z} is adapted by replacing $z_{\textrm{30}}$ with $z_{\textrm{SOM}}$:
\begin{equation}\label{eq:som-outliers}
|z_{\mathrm{SOM}} - z_{\mathrm{spec}}| >  0.15(1+z_{\mathrm{spec}}).
\end{equation}

We draw similar conclusions to those summarized in Table~\ref{tab:match}: the blended samples have significantly higher photo-\textit{z} outlier rates than the pure samples; the 7-feature sample has more outliers than the 10-feature sample; the 10-feature strong blends sample has a higher outlier rate than the 10-feature all blends sample. 
However, the 7-feature strong blend sample has a lower outlier rate than the 7-feature all blend sample.
We believe this is a statistical fluctuation due to a small sample size.
\begin{table}
	\caption{Summary of the validation sample and its sub-samples for SOM\textit{z} with 10 or 7 features, labeled 10-SOM\textit{z} and 7-SOM\textit{z}, respectively. The validation sample consists of half of the pure sample and half of the blending sample, as defined in Sect.~\ref{subsec:match}. The SOM\textit{z} outliers in this table are defined by $|z_{\mathrm{SOM}} - z_{\mathrm{spec}}| >  0.15(1+z_{\mathrm{spec}})$, and the outlier fraction is the fraction of photo-\textit{z} outliers over the spec-match subsamples.}
	\label{tab:somz}
	\centering
	\begin{tabular}{rcccc}
		\toprule
                                           &  Total   &   Pure   &   All BL   & Strong BL  \\
		Num. Sources                       &  72\,317   &   61\,070  &    11\,247   &   2\,690     \\
		Pure Fraction                      &  84.5\,\% &   100\,\%  &    0\,\%     &    0\,\%     \\
		Num. Spectra Match                 &  20\,078   &   18\,370  &    1\,708    &   419      \\
		Num. 10-SOM\textit{z} Outliers     &   804    &    633   &    171     &   48       \\
		Frac. 10-SOM\textit{z} Outliers    &  4.0\,\%  &   3.5\,\% &   10\,\%  &  12\,\%   \\
  		Num. 7-SOM\textit{z} Outliers      &   1\,048   &    805   &    243     &   55       \\
		Frac. 7-SOM\textit{z} Outliers     &  5.2\,\%  &   4.5\,\% &   14\,\%  &  13\,\%   \\
        \bottomrule
    \end{tabular}
\end{table}

%% file: 4.0.tex
\section{Results}
\label{sec:res}


We use a variety of machine learning algorithms to label blends in the validation sample.
All algorithms, aside from One-Class SVM, assign a score to objects, which allows us to rank objects from likely pure galaxies to likely blended objects.
For RF, k-NN, and anomaly detections, objects are ranked by the score from that method, with high scores corresponding to blends and low score to pure sources.
For the SOM blending cell approach, the score of an object is the blend-to-pure occupancy ratio of the cell it maps onto.
For the SOM distance cut, the score is simply the $\chi^2$ distance to its best-match cell, where a larger distance indicates a more likely blend.
In the following sections, ``SOM'' will refer to the blending cell approach unless otherwise specified.
While we are only classifying objects, the typical use case would be to remove unrecognized blends from a sample, so we use the term removing and labelling interchangeably.




Our results on detecting unrecognized blends are presented in terms of two parameters: \textbf{Recall}, the fraction of blends labeled as blends, and \textbf{Cost}, the fraction of all detections (blended or not) labeled as blends, \edit{as defined below in Eq.~\ref{eq:recall_cost}.
Let $B$ be the total number of unrecognized blends in the validation sample, $P$ the total number of pure galaxies, $R_b$ be the number of removed unrecognized blends, and $R_p$ the number of removed pure galaxies. We define
\begin{equation}
\begin{split}\label{eq:recall_cost}
    &\textrm{Cost} = \frac{R_b+R_p}{B+P}; \;\;\;\; \textrm{Recall} = \frac{R_b}{B};\end{split}    
\end{equation}
These metrics were chosen to highlight our initial goal, to improve sample purity and quality and understanding the cost to the sample we would incur.
Unlike other classification projects, we are not currently interested in detailed study of the objects we are classifying (unrecognized blends) but hope that we can use them to improve sample purity at the cost of statistical power which recall and cost help probe. }

There may be interest in other classification metrics such as precision and remaining blend fraction -- precise definitions and performance with these other metrics are shown in App.~\ref{sec:app-class}. 

In the following, Sect.~\ref{subsec:rm} shows the main results of blends detection from various algorithms.
In Sect.~\ref{subsec:ws_blends}, we test the performance of the algorithms for removing ``strong blends'', a more stringent blending sample defined in Sect.~\ref{subsec:match}.
In Sect.~\ref{subsec:alt_conf}, we explore a variety of different configurations for SOM and Random Forest and in Sect.~\ref{subsec:pz_res} we investigate how removing unrecognized blends can be used to indirectly remove photo-z outliers.
We conclude our results by studying the importance of various catalog features in Sect.~\ref{subsec:dis_imp}.


%% file: 4.1.tex
\subsection{Removing Unrecognized Blends}
\label{subsec:rm}

\begin{figure*}
    \centering
    \includegraphics[width=0.8\linewidth]{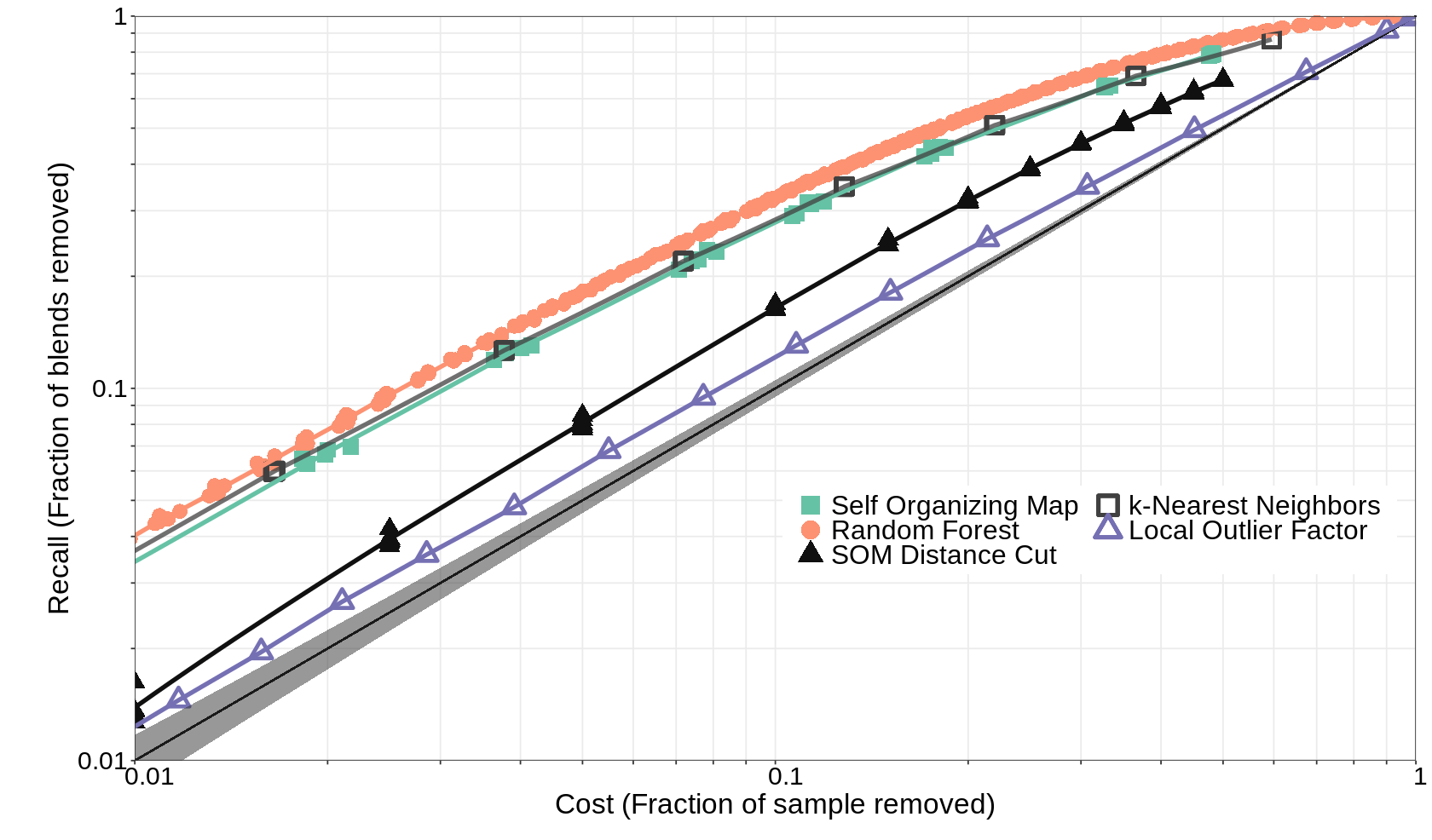}
    \caption{Main results of this work. Removing potential blended sources from the validation sample using different algorithms and quantifying the performance in terms of Recall, the fraction of blends removed, versus Cost, the fraction of all detected sources removed, both defined in Eq.~\ref{eq:precision_recall}. SOM results are shown in green squares, RF in orange circles, LOF in purple triangles, and k-NN in hollow black squares. We include a baseline curve in gray representing a random, or no-skill, classifier along with a 95\% confidence interval derived from a hypergeometric distribution as the shaded region around the baseline. One anomaly detection method, Local Outlier Factor, is plotted as a comparison. It has the best results out of the 4 anomaly detection methods considered in this work, yet it behaves only marginally better than a random classification, while RF, SOM, and k-NN have substantial better performances. Selected results for SOM and RF are shown in Table~\ref{tab:res}.}
    \label{fig:res}
\end{figure*}

With the SOM blending cell approach, we remove galaxies from the validation sample that are mapped onto the top 10, 50, 100, 200, 300, 500, 1\,000, or 1\,500 blending cells, ranked by the ratio of blends to pure galaxies from high to low. 
This roughly corresponds to removing 1\,\% to 50\,\% of the total sample.
For the SOM distance cut method, we remove 1\,\% to 50\,\% detections according to their distance to best-match cells. 
For RF, we remove 1\,\% to 100\,\% detections by requiring a score greater than some varying threshold, and the same with k-NN. We repeat this process for 5 different realizations using different training-validation data splits for RF and SOM.
Our main results are presented in Fig.~\ref{fig:res} and Table~\ref{tab:res}. For comparison, we show results from Local Outlier Factor, the best-performance anomaly detection out of 4 algorithms tested in this work (Sect.~\ref{subsec:knn}). Other anomaly detection algorithms are shown in App.~\ref{sec:anom}.


Results shown in this section are based on \textbf{all} blends (Sect.~\ref{subsec:match}).
As a reminder, the all blends sample are objects with a magnitude difference $\Delta_i \leq 2$ between the ground detection and its second HST component. At a moderate Cost of 10\,\% of all detected sources, likely an acceptable cut for most applications, the SOM blending cell approach identifies 28\,\% of all unrecognized blends and the distance cut identifies only 18\,\%. RF has the best performance identifying 33\,\% of blends.
For applications that prioritize sample purity, we are able to remove 90\,\% of blends at a cost of 60\,\% of all detected sources when using SOM blending cell or 55\,\% when using RF.

\begin{figure*}
    \centering
    \includegraphics[width=1.0\linewidth]{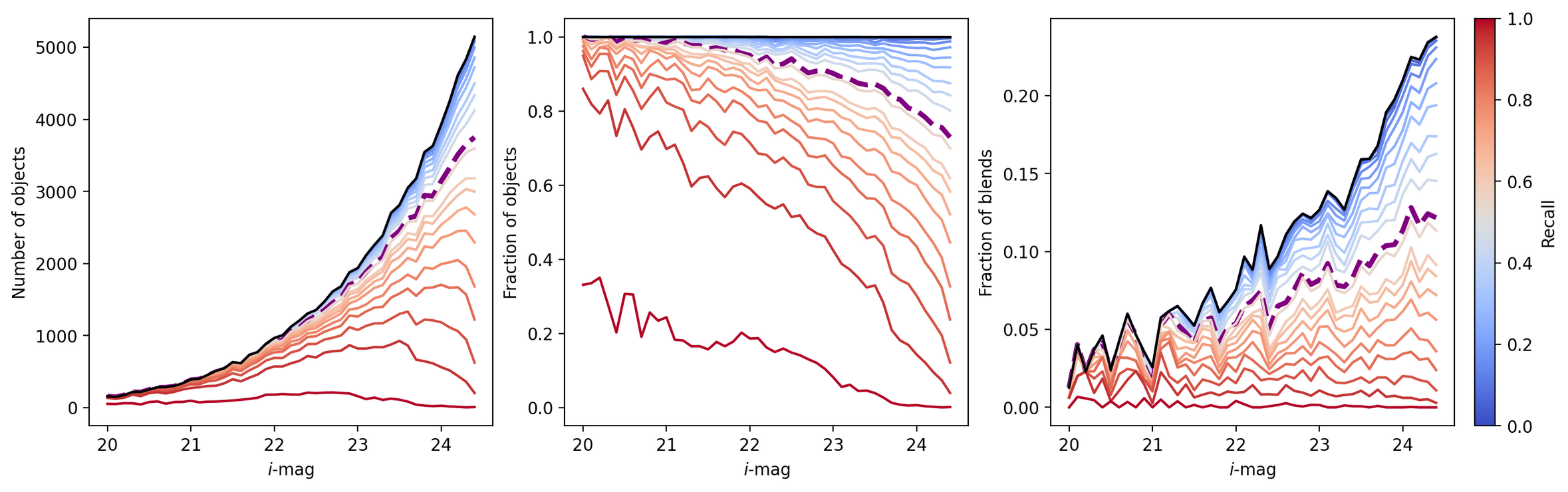}
    \caption{\edit{$i$-mag distributions with varying Recall values using bins of width 0.1 mags from 20 to 24.5. Removed objects include unrecognized blends and pure objects.  The distributions correspond to a range of Recall values between 0 (blue) and 1 (red) with a step size of 0.05. 50\,\% Recall is shown as a bold purple dashed line and the original sample distribution is shown as a solid black line. \textbf{Left}: \textit{i}-mag distribution of the sample after removing possible blends. \textbf{Middle}: Fraction of objects that remain after removing possible blends per $i$-mag bin. The most aggressive cut leaves 40\% of the objects with $i$-mag $= 20$. \textbf{Right}: Blend fraction per $i$-mag bin.
    }}
    \label{fig:ibin-changes}
\end{figure*}

At high Recall values, we are able to remove the majority of blends but at the expense of removing large fractions of pure objects.
\edit{We plot the \textit{i}-mag distribution and fraction of blends per \textit{i}-mag bin as a function of recall in Fig.~\ref{fig:ibin-changes}.
On the left we can see that as Recall increases, the total number of galaxies drops to 0 as the algorithm labels all objects as blends.
The middle panel normalizes the left plot with the original counts per \textit{i}-mag bin (recall line divided by black line) showing that the main effect is removing faint objects while there are still some bright objects even at high Recall.
On the right plot, this \textit{i}-mag dependence is supported once again as the blend fraction decreases the most dramatically for faint objects (\textit{i}-mag > 22) hinting at the importance of the \textit{i}-band which we discuss in greater detail in Sect.~\ref{subsec:dis_imp}.}

\begin{table}
\centering
\caption{Summary statistics of the SOM and RF blend detection. The left half shows the results for removing SOM cells with high blends-to-pure occupancy ratio and the right half for removing high-score blending objects as predicted by Random Forest. Cell refers to the number of cells removed from the SOM (see Sect.~\ref{subsec:conf}), Cost is the fraction of all detections labeled as a blend, and Recall is the fraction of the total blends labeled as blends. Precise definitions and the results using other summary statistics can be found in App.~\ref{sec:app-class}. The results shown are averaged over 5 trainings. All numbers are in percentiles except for the Cell number (first column). Results from the 5 realizations are plotted in Fig.~\ref{fig:res}.}
\begin{tabular}{RRRRR}
\toprule
\multicolumn{3}{c}{SOM Blending Cell}  &  \multicolumn{2}{c}{Random Forest} \\
\cmidrule(lr){1-3}
\cmidrule(lr){4-5}

\multicolumn{1}{r}{Cell}       &
\multicolumn{1}{r}{Cost}       &
\multicolumn{1}{r}{Recall}     &
\multicolumn{1}{r}{Cost}       &
\multicolumn{1}{r}{Recall}     \\ \midrule

 10   & 0.77  & 2.64   &  0.80  &  3.29  \\
 50   & 3.16  & 9.97   &  3.19  & 11.98  \\
 100  & 5.90  & 17.21  &   5.45 & 19.22  \\
 200  & 10.58 & 28.02  &  10.60 & 33.63  \\
 300  & 14.61 & 36.14  &  14.67 & 42.98  \\
 500  & 22.01 & 48.58  &  22.71 & 57.83  \\
 1\,000 & 38.20 & 68.56  &  38.47 & 77.14  \\
 1\,500 & 53.16 & 80.96  &  54.57 & 89.35  \\
 \bottomrule \\
\end{tabular}
\label{tab:res}
\end{table}

%% file: 4.2.tex
\subsection{All Blends vs. Strong Blends}
\label{subsec:ws_blends}

We further test the performance of SOM and RF on a more stringent blends sample, the strong blends (Sect.~\ref{subsec:match}): unrecognized blends with a magnitude difference $\Delta_i \leq 1$ between the ground detection and its second HST component. 
With a smaller difference in magnitude, the second component yields a greater impact on flux and shape measurements of strong blends, which motivates prioritizing the detection of them. 
By definition, strong blends will likely have more exotic colors and are expected to be identified more easily with machine learning algorithms.
As seen in Table~\ref{tab:match}, strong blends are also more likely to be photo-\textit{z} outliers.

\begin{figure}
	\centering
            \includegraphics[width=1\columnwidth]{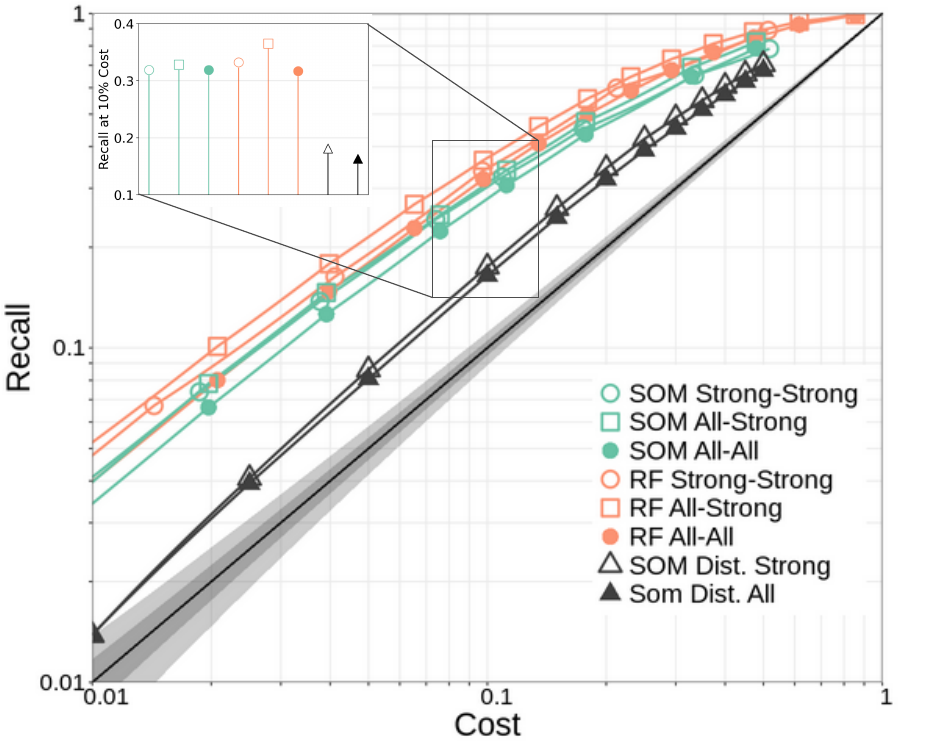}
	\caption{\edit{Detecting ``all'' and ``strong'' blends as defined in Sect.~\ref{subsec:ws_blends}. An inset is included using the same symbols as the larger figure to show the Recall values at a fixed 10\% cost. Both SOM and RF perform the best using all blends to predict strong blends (All-Strong). The distance cut (Sect.~\ref{subsec:conf}) is an alternative method for identifying blends with SOM rather than the fiducial ``blending cell'' method. This method does not have a All-All/All-Strong/Strong-Strong distinction since it only needs the pure training information coded into SOM. The All-All/Strong-Strong points of SOM Distance simply represent detecting weak/strong blends with a distance cut. All plotted points are averaged over 5 trainings.}} 
	\label{fig:strong_dist}
\end{figure}

We vary the type of blends in a few scenarios. For the SOM blending cell approach, the training set stays the same pure training sample as defined in Sect.~\ref{subsec:match}, while the blends definition in the identification and validation sample varies. For RF, both the training and validation samples change.
We focus on three cases:
\begin{enumerate}[leftmargin=*,labelindent=10pt,label=\arabic*.]
    \item All-All (A-A): Use \textbf{all} blends to identify blending cells for SOM, and quantify the removal of \textbf{all} blends. For RF, use pure + \textbf{all} blends as the training sample, and quantify \textbf{all} blends in the validation sample. This corresponds to the fiducial results presented in Sect.~\ref{subsec:rm}.
    \item All-Strong (A-S): Use \textbf{all} blends to identify blending cells for SOM, and quantify the removal of \textbf{strong} blends. For RF, use pure + \textbf{all} blends as the training sample, and quantify \textbf{strong} blends in the validation sample.
    \item Strong-Strong (S-S): Use \textbf{strong} blends to identify blending cells for SOM, and quantify the removal of \textbf{strong} blends. For RF, use pure + \textbf{all} blends as the training sample, but labelling the pure + weak blends as \textbf{pure}, and \textit{\textbf{only}} the strong blends as blends. Weak blends are included in training however they are not labeled as blends. Then quantify the \textbf{strong} blends in the validation sample.
\end{enumerate}
The SOM distance approach does not need the identification of blending cells, thus we consider two scenarios:
\begin{enumerate}[leftmargin=*,labelindent=10pt,label=\arabic*.]
    \item All (A): train the SOM with the pure sample, and quantify the removal of \textbf{all} blends according to their $\chi^2$ distances.
    \item Strong (S): train the SOM with the pure sample, and quantify the removal of \textbf{strong} blends according to their $\chi^2$ distances.
\end{enumerate}
Results for all of the above scenarios are plotted in Fig.~\ref{fig:strong_dist}.
All three algorithms are consistent among the cases with slightly improved performance in the order of A-A < S-S < A-S or A < S. This means: 
\begin{enumerate}[leftmargin=*,labelindent=10pt,label=\arabic*.]
\item Strong blends are easier to identify than all blends (A-A < S-S and A < S). This is likely due to their colors being more different from pure galaxies. 
\item Adding weak blends helps with finding strong blends (S-S < A-S). This indicates that strong blends occupy the same feature spaces as weak blends. The addition of weak blends provides more samples for identifying blending cells, and thus helps with the performance by increasing the sensitivity of the training.
\end{enumerate}

Nearest neighbors supports the same conclusions of A-A < S-S < A-S with the Recall values being $33.4\%$, $35.2\%$, and $41.2\%$, respectively, at a $10\%$ Cost.

%% file: 4.3.tex
\subsection{Alternative Configurations}
\label{subsec:alt_conf}
\begin{figure}
    \centering
        \includegraphics[width=\columnwidth]{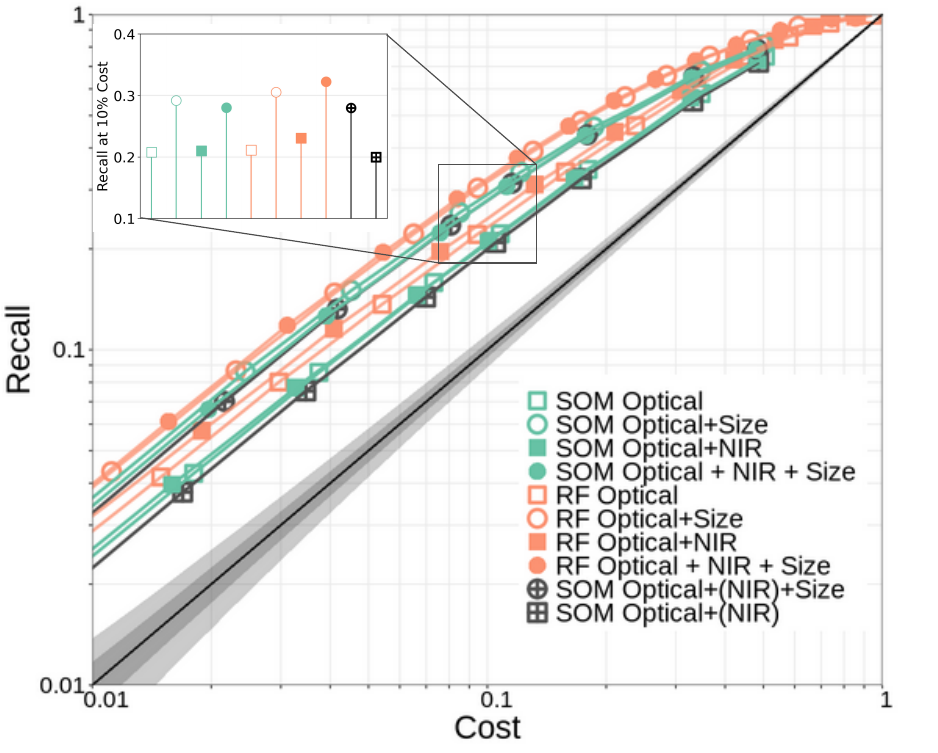}
    \caption{\edit{A selection of alternative configurations for SOM and RF as discussed in Sect.~\ref{subsec:ws_blends}. Data points plotted using a circle (solid or hollow) use the size parameter, square without size. Solid points (circle or square) use NIR bands, hollow points do not use NIR, and the half-filled black points use NIR for training but not for validation --- notated in the legend using parentheses around ``NIR''. An inset is included using the same symbols as the larger figure to show the values at a fixed 10\% cost. The results shown are based on an All-All set up. We find that size is more important in detecting blends than the infrared bands. More configuration set-ups are discussed in App.~\ref{sec:feat_varia}.}}
    \label{fig:feat_conf}
\end{figure}

We further explore a variety of alternative configurations for detecting blends focusing on the future synergies with NIR joint datasets.
Selected configurations are listed below with the rest presented in App.~\ref{sec:feat_varia}.
We consider:
\begin{enumerate}[leftmargin=*,labelindent=10pt,label=\arabic*.]
    \item 10+10, the fiducial run: using 10 features (8 colors, $i^{+}$ magnitude, size) for training/identification, and the same 10 features for validation.
    \item 9+9, no size: using 9 features (8 colors, $i^{+}$ magnitude) for all steps: training/identification/validation.
    \item 7+7, optical and size: using optical bands and size (5 colors, $i^{+}$ magnitude, size) for all steps.
    \item 6+6, optical only: using optical photometry only (no size) for all steps.
    \item 10+7: using 10 features for training and identification, but only 7 features (optical bands and size) for validation. This mimics the scenario where a small region of full band coverage is available for training.
    \item 9+6: same as the above case, but not using size.
\end{enumerate}
We show results using an All-All set-up (no differentiation between the weak and strong blends) in Fig.~\ref{fig:feat_conf}, and we omit the SOM distance cut results since it underperforms the other two methods.
Our main conclusion is that dropping the near-infrared bands has almost no impact on the results, while dropping the size information has a very significant impact on Recall. The size alone affects the Recall by nearly 0.1 at the 10\,\% cost. This means that our method will perform equally well on the Rubin-LSST data set with or without any additional near-infrared data. We also find that for the 10-feature SOM, dropping the near-infrared bands actually improved the results slightly (7+7 is better than 10+10).
\edit{The correlation between different bands along with the lower dimensionality using only optical data may make it easier for the 7-feature SOM to cleanly identify blended regions.}
A more detailed analysis of the importance of features can be found in Sec.~\ref{subsec:dis_imp}, and discussion on more alternative configurations can be found in App.~\ref{sec:feat_varia}.

\edit{We note that X-ray AGNs were removed from the labeled sample which may influence the conclusions of the results shown in this section.
The conclusion that NIR bands are not necessary to detect unrecognized blends does not meaningfully change by including AGNs in our sample as only $\sim 200$ of the AGNs were labelled as unrecognized blends.
However it is possible that a multi-class classifier that properly takes AGNs into account, along with stars, galaxies, and blends, will almost certainly perform better by including NIR bands.
This is out of the scope of this work but the authors look forward to future studies on this.}

%% file: 4.4.tex
\subsection{Photo-\textit{z} Outliers}
\label{subsec:pz_res}

By removing a fraction of unrecognized blends, we expect the remaining sample to have more accurate colors, shapes, and photometric redshifts.
Based on the expectation that blended sources tend to be photo-\textit{z} outliers (as seen in Table~\ref{tab:somz}), we predict that the ``feature outliers'', whose high-dimensional feature (colors+size) are the largest offenders compared to pure galaxies, will likely be both blended objects and photo-$z$ outliers. 
Therefore, we quantify photo-$z$ outliers in the validation sample after removing predicted blends. Here we define photo-\textit{z} outliers with $z_{\mathrm{SOM}}$, as defined in Eq.~\ref{eq:som-outliers}.
We omit the 7-band SOM$z$ results since they are almost identical to the 10-band SOM$z$ results.
Note that the definition of photo-$z$ outliers is only valid for the sub-sample of objects with spectroscopic redshifts.
We study a variety of configurations for photo-$z$ outliers:
\begin{enumerate}[leftmargin=*,labelindent=10pt,label=\arabic*.]
    \item SOM All: remove objects from blending cells identified with All blends;
    \item SOM Distance Cut: remove objects using the $\chi^2$ distance cut;
    \item RF All: train the RF with all blends labelled as blends;
    \item Anomaly detection: all methods listed in Sect.~\ref{subsec:anomaly}.
\end{enumerate}

\begin{figure*}
    \centering
    \includegraphics[width=0.48\linewidth]{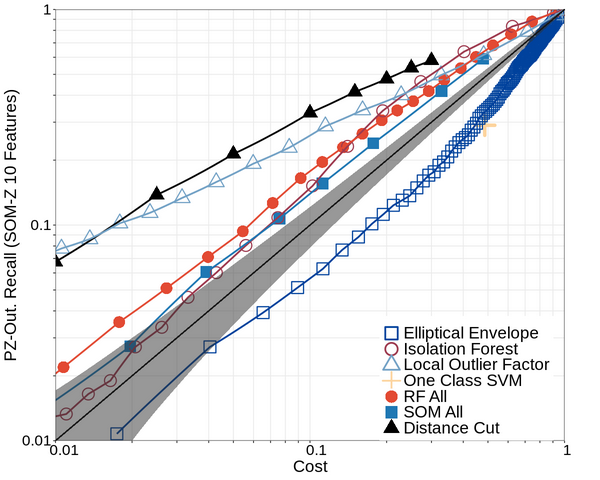}
    \includegraphics[width=0.48\linewidth]{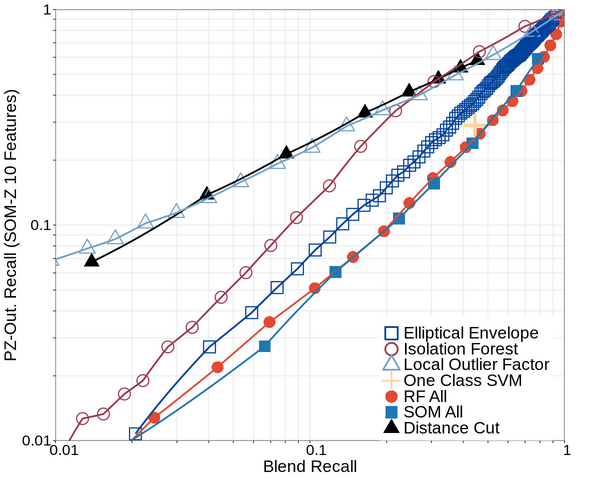}
    \caption{Summary statistics on removing photo-\textit{z} outliers by removing blends. The fraction of photo-\textit{z} outliers removed from all photo-\textit{z} outliers (labeled as PZ-out. recall) is plotted against the cost to the all detections on the left and the fraction of blends correctly labeled as blends (blend recall) on the right.  The photo-\textit{z}s are generated using a 10 feature SOM as outlined in Sect.~\ref{subsec:somz}. The SOM results are shown in solid blue squares, SOM distance cut results in solid black triangles, the RF results in solid red circles, and all anomaly detection methods in hollow markers. The SOM distance cut and LOF deliver significantly better results than RF and the SOM blending approach. This reveals that the unrecognized blends and photo-\textit{z} outliers are different populations, which reside at different locations in the feature space.}
    \label{fig:pz_outlier}
\end{figure*}

In the left panel of Fig.~\ref{fig:pz_outlier} we show the Recall on photo-$z$ outliers as a function of Cost to all detected sources.
We find that SOM-All and RF-All are only marginally better than a random selection of objects as indicated by the diagonal line.
Surprisingly, the SOM distance cut method outperforms the previous methods at all costs, with LOF performing only slightly worse.
Isolation Forest shows a performance consistent with random removal at low cost but offers high efficiency at Cost $> 0.4$.
The other anomaly detection methods do not perform well in this comparison, with EE and One-Class SVM performing worse than a random selection. 

In the right panel of Fig.~\ref{fig:pz_outlier} we present the Recall on photo-$z$ outliers versus the Recall on All blends.
This direct comparison shows how effective our blend-oriented algorithms are at identifying photo-\textit{z} outliers.
Again LOF and the SOM distance cut perform significantly better than the others, while the SOM blending cell approach and RF are among the worst, underperforming even the simple anomaly detection methods. We do not compare the results with a one-to-one diagonal line in this panel as the quantitative relation between detected unrecognized blends and photo-\textit{z} outliers can be complicated. 

It is important to note that our implementation of RF and the SOM blending cell approach are \edit{designed} to detect unrecognized blends, while the SOM distance cut and LOF are set up to detect ``feature outliers''. 
\edit{RF is an explicitly supervised algorithm to label regions of feature space while SOM is an unsupervised reduction of that feature space to a 2D map.
However, by identifying regions on the SOM that correspond to high blend fractions, we have introduced labeling which makes it better suited for detecting unrecognized blends.
This is contrasted with SOM distance cut and LOF which have no labeling at any point and are closer to traditional ``unsupervised'' algorithms.
}
The fact that the ``feature outlier'' approaches outperform the previous two indicates that the photo-$z$ outliers have more exotic features than blends. 
This point is strengthened by LOF, which is aimed at finding isolated objects in feature space where a low local density is found. 
In other words, LOF is prioritized to find the most extreme features. On the other hand, this means that the blended objects are less extreme and often share their feature spaces with pure galaxies.
Those features spaces can only be identified by targeted Machine Learning algorithms such as RF and the SOM blending cell approach. We discuss the relative importance of each feature for identifying such feature spaces in Sect.~\ref{subsec:dis_imp}.
In conclusion, unrecognized blends and photo-$z$ outliers are two dissimilar populations based on the features considered in this work. The SOM distance cut and LOF are efficient at detecting the ``feature outliers'', of which a good proportion are photo-$z$ outliers.
At a 10\,\% cost to the sample, the SOM distance cut is able to remove 18\,\% of unrecognized blends which includes 35\,\% of the total photo-$z$ outliers.

A more direct method to detect photo-\textit{z} outliers would be to train the RF with a photo-z ``inlier + outlier'' sample instead of the pure + blends sample, or to map an ``outlier identification sample'' onto the SOM to find "outlier cells". We choose not to explore this set-up for two reasons. First, we have a limited sample size  -- only 804 photo-\textit{z} outliers in total, and we prefer to reserve all of them for validation. Second, as shown in Fig.~\ref{fig:mag_hist}, the spectrum-matched subsample underwent a very different selection from the validation sample. Training and validating on such a subsample does not necessarily imply any useful information about real-world applications. We seek to study the problem more thoroughly in the future with both larger samples of real observations and simulated data.

We emphasize again that the definition of photo-\textit{z} outliers is ambiguous for blended objects since there is no single true redshift. We look forward to further studies on disentangling photo-\textit{z} with blends.

%% file: 4.5.tex
\subsection{Importance of Features}
\label{subsec:dis_imp}

\begin{figure}
    \centering
    \includegraphics[width=1\linewidth]{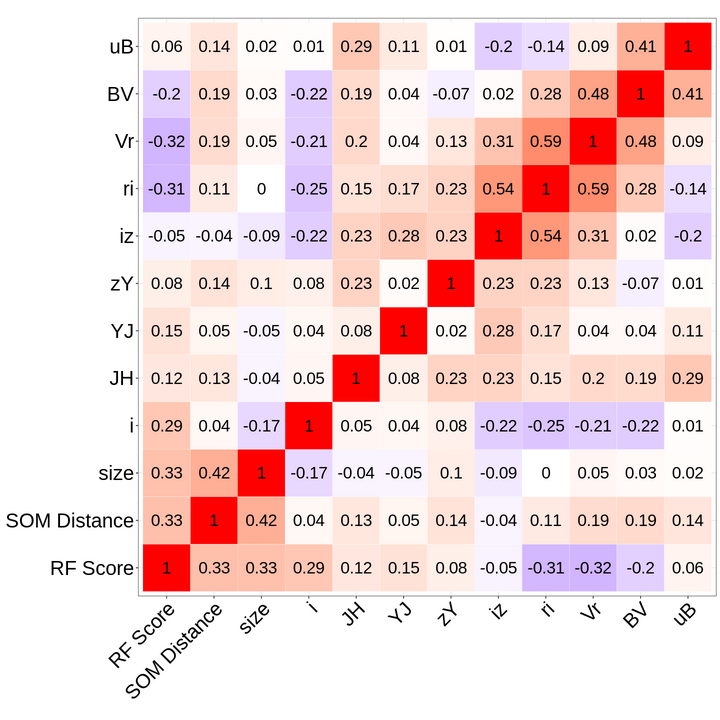}
    \cprotect\caption{Correlation matrix using \verb|score|$_{\textrm{RF}}$ from RF and \verb|distance| from SOM distance, and all input features colored with dark blue being a correlation of -1 and dark red being a correlation of 1.}
    \label{fig:correlation}
\end{figure}

\begin{figure}
    \centering
    \includegraphics[width=\linewidth]{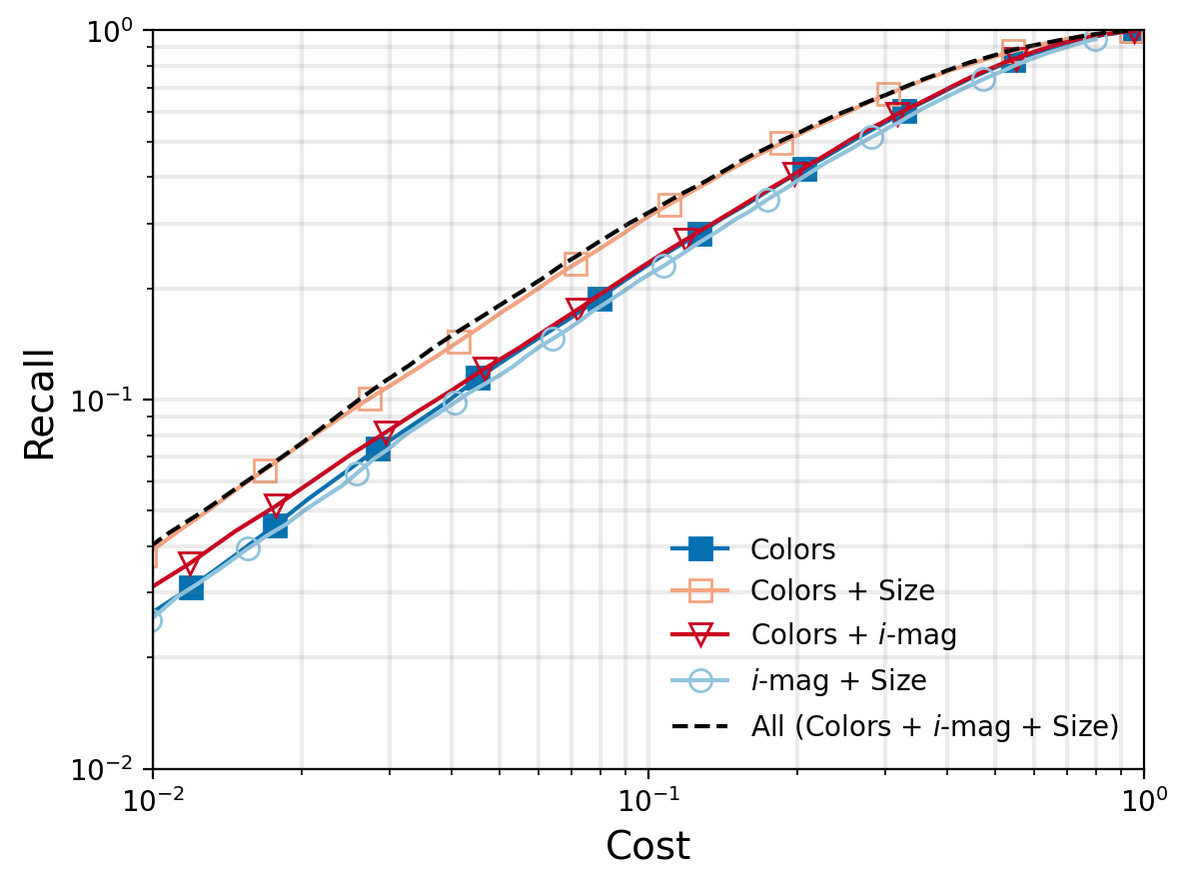}
    \caption{Select configurations of training features for RF to highlight the importance of color information.  In black we include the performance of RF when trained on all features available as a benchmark. While $i$-magnitude and size measurements are very important as shown later in Sect.~\ref{subsec:dis_imp}, they perform worse than using all optical and NIR colors. Excluding the $i$-magnitude but including all color information does nearly identically to the all feature benchmark. Note that while we re-use markers from previous graphs, all the results in this figure are focused on RF only.}
    \label{fig:RF-subsets}
\end{figure}

As discussed in Sect.~\ref{subsec:alt_conf} and illustrated in Fig.~\ref{fig:feat_conf}, removing the \verb|size| feature has a large impact on performance whereas removing the 3 infrared bands has a smaller effect.
In Fig.~\ref{fig:ibin-changes} we see the non-uniform impact of \textit{i}-mag on removing unrecognized blends.
In this section we further investigate the importance of various features. 
We start by calculating the correlation coefficient between the features and our predictors -- the RF \texttt{score} and the SOM \texttt{distance}, shown in Fig.~\ref{fig:correlation}.
\edit{We use SOM \texttt{distance} as opposed to the blending cell ratio since this cleanly gives each object a score whereas the blending cell ratio gives areas of color-space the same score which makes correlation calculations non-trivial.}
We also show a variation of Fig.~\ref{fig:feat_conf}, using RF only, to compare the relative importance among colors (Optical + NIR), $i$-magnitude, and size measurement, in Fig.~\ref{fig:RF-subsets}.

From the correlations we see that size is a strong predictor for SOM \verb|distance| and RF \verb|score|.
RF \verb|score| is also strongly correlated with \textit{i}-band magnitude and colors \textit{r-i} and \textit{V-r}, while SOM \verb|distance| is only weakly correlated with the colors and almost uncorrelated with \textit{i}-band magnitude.
\edit{The low correlation between SOM \texttt{distance} and \textit{i}-band magnitude could explain the low performance of the classifier (note that \textit{i}-band was used as a feature in the SOM). 
}
In Fig.~\ref{fig:RF-subsets}, we show the impact of changing the features that are included. Using all colors is marginally better than using $i$-mag and a size measurement, while using a combination of colors and size measurement is able to match the performance of all features for all Costs.
These two cursory methods of understanding the importance of features motivates the use of a more robust method of estimating variable importance, Recursive Feature Elimination \citep[RFE;][]{rfe_cancer}. 




Given a metric to characterize performance, RFE ranks variables by re-training models with a single feature missing, measuring the performance metric, and ranking based on which feature changes the chosen metric the least.
That feature is then removed for subsequent runs and the process is repeated until all features are ranked.
The metric chosen is the Recall at Cost$=0.1$ which is $0.32$ in the fiducial case using all features.
When removing a band, all remaining bands are re-combined to create colors; however, the $i$-band magnitude remains as a single magnitude measurement. 

\begin{figure*}
    \centering
    \includegraphics[width=\columnwidth]{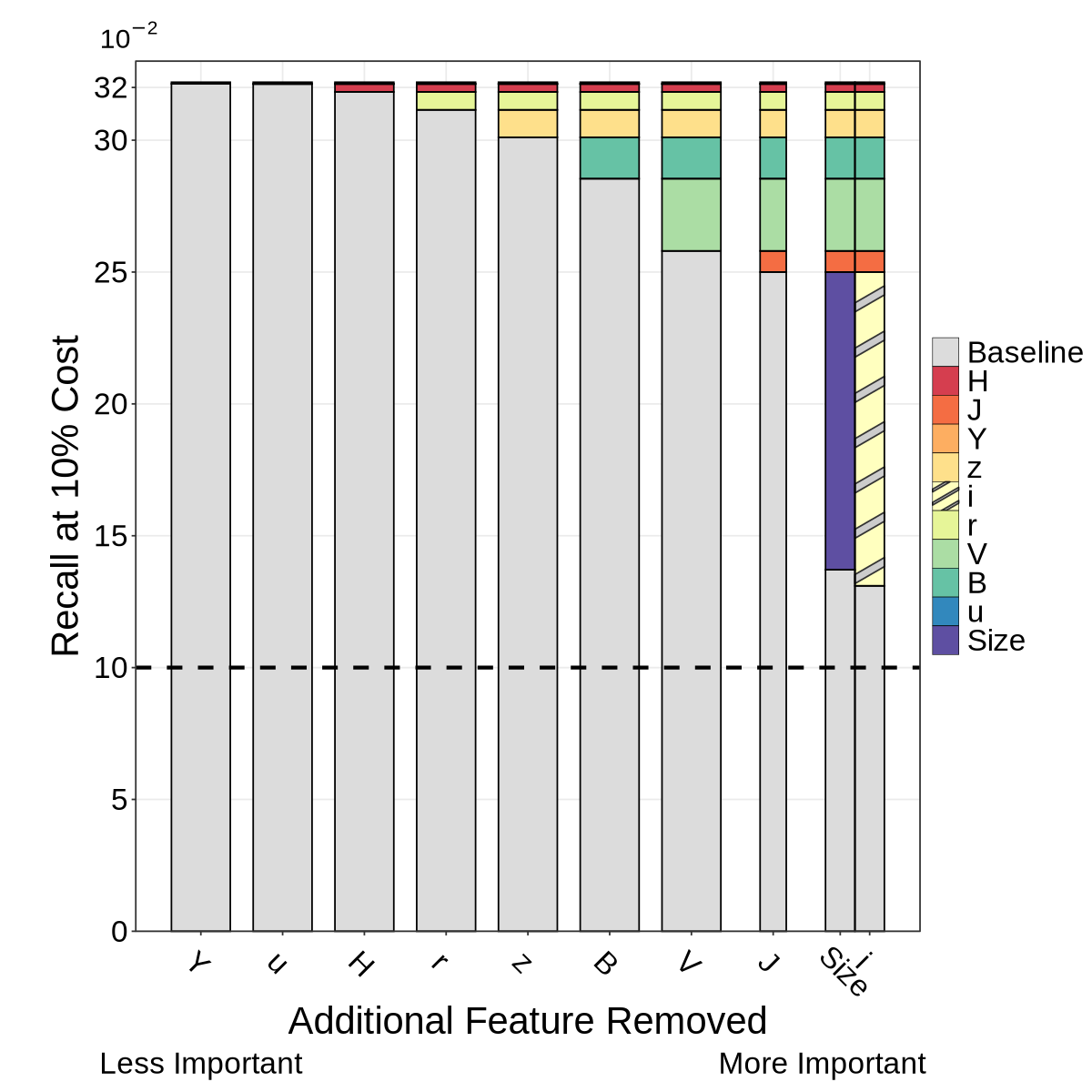}
    \includegraphics[width=\columnwidth]{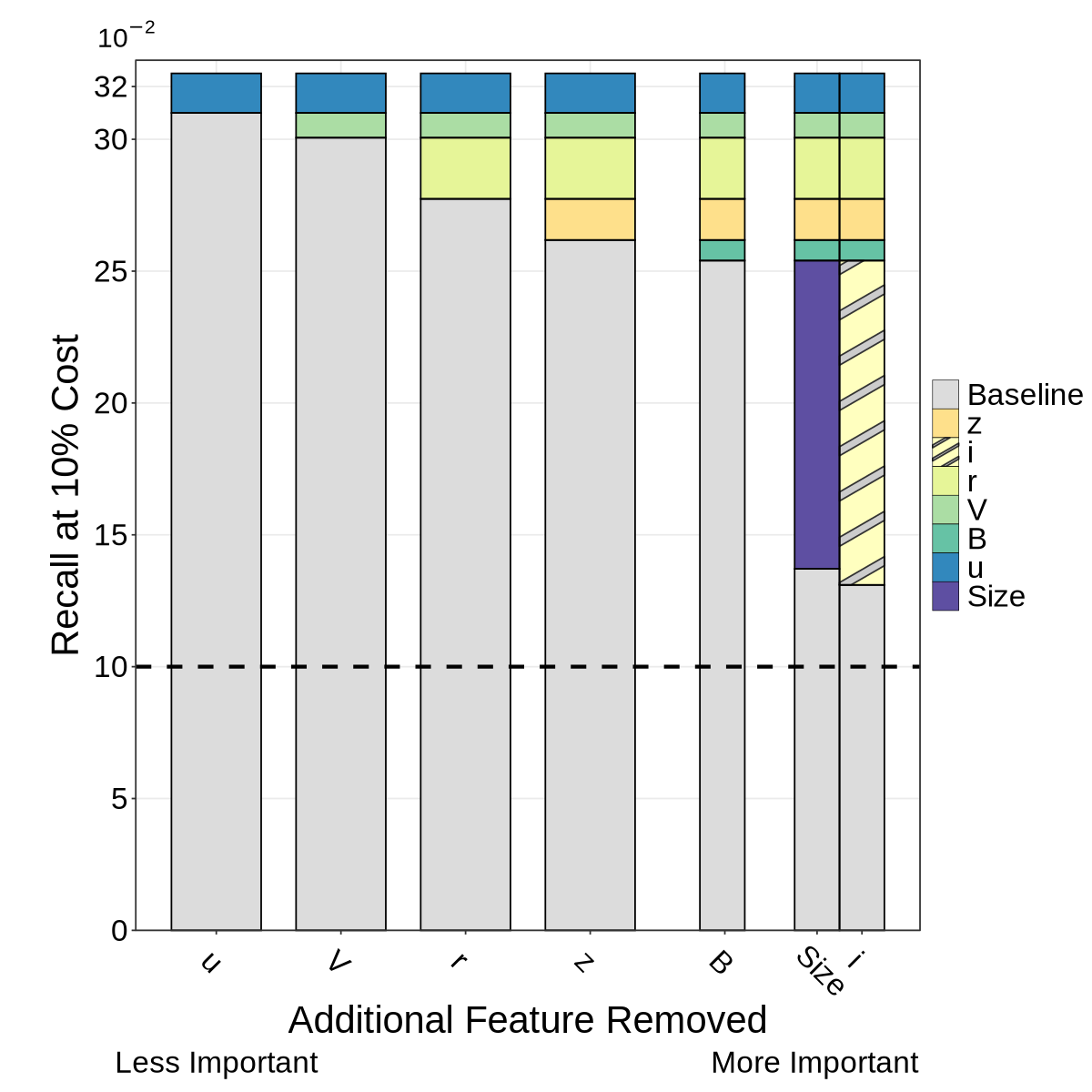}
    \caption{\edit{Importance of features ranked by decrease in performance of RF to identify unrecognized blends at 10\% Cost of total sample. The horizontal axis names features in increasing importance from left to right and the vertical axis displays the recall at a 10\% Cost to the sample when that feature is removed. The loss in recall by removing a feature is displayed in subsequent runs with the color-coded rectangles. A dashed line at 10\% Recall is included to indicate the performance of a random selection of objects at 10\% Cost. Features are organized from least to most impact from left to right. When possible, remaining bands are re-combined into colors and the $i^{\p}$-band is left as magnitude. Full width bars are used when the forests are trained on colors while half-width uses magnitudes only. The last two bars are trained on one feature only, either only $i^{\p}$-band or only size. Left: Using all features including the optical and the NIR bands as well as the size. Right: Using only the optical bands and the size.}}
    \label{fig:feature_importance}
\end{figure*}

Results are shown in Fig.~\ref{fig:feature_importance} with the horizontal axis labeling the least to most important features from left to right. 
The plot on the left presents the results when using all bands available, optical and NIR, while the right plot focuses on optical bands only. 
When there are only two magnitudes left, it becomes impossible to iterate by removing another feature and still create colors so instead we train on two magnitudes.
This is reflected in the plots by including a gap between the forests trained on colors + magnitude + size versus magnitudes + size.
At this point we no longer enforce that the \textit{i}-mag remain as a single magnitude measurement and can be removed.
After removing the 3rd most important feature (\textit{J}-mag on left and \textit{B}-mag for right) from the forests, there are two features left, size and \textit{i}-mag, which means after removing a feature, the forest can only be trained on one feature, which is why the last two columns are presented bordering to each. 
We find consistent results, in terms of order of bands, when changing which band is fixed as the single magnitude.

In both setups, testing on optical only and optical+NIR bands, we find that \verb|size| and \textit{i}-mag are the most important features.
The \textit{i}-band is deep in both the COSMOS and HST data which can explain its importance.
The \verb|size| feature is expected since it is likely that blends will have a larger angular size than pure galaxies, and as seen in Fig.~\ref{fig:correlation}, the size is uncorrelated with colors and therefore provides unique information content.
The \textit{u}-band has substantial overlap with the \textit{B}-band so it is possible that if the \textit{B}-band were not included, \textit{u} would become more informative but with our study, the \textit{u}-band is one of the least informative features.
Oddly, the \textit{J}-band is ranked highly even if the exclusion of NIR bands leads to a negligible decrease in performance Fig.~\ref{fig:feat_conf}.
The interplay between bands is non-linear and may increase the importance of the \textit{J}-band after removing the other NIR bands.
While the RFE does give us a ranking of features, it is important to note that with optical data every band has a noticeable effect on performance.

\edit{The importance of the size feature hints that other shape measurements may offer even more classification power, e.g. ellipticity. However, ellipticity must be treated with care, since a selection on ellipticity will likely induce a shear-dependent bias which we would like to avoid for weak lensing cosmology. It is possible that the use of the size measurement itself for blends removal will create a bias in cosmic shear already. However, as this study shows, in the worst case we are still able to identify unrecognized blends with colors alone. 
On-going work \citep{rf_anacal} suggests that the selection bias from including shape information cannot be controlled but size information can.
}

%% file: 5.2.tex
\section{Implications for Cosmology}
\label{subsec:dis_app}
One potential application of this work is cosmology from galaxy surveys, where correlations between galaxy positions and shapes (usually termed as ``$3\times2$pt correlation'') are used to probe large scale structure in the universe. 
Unrecognized blends have a major impact on photometric redshift and shear measurements, both of which are needed for $3\times2$pt cosmology.
This is exacerbated by the fact that the existence of unrecognized blends correlates shear measurements with photometric redshifts. 
As explained in \citet{des_bl}, this can be understood by thinking of the measured shear at one position on the sky as a weighted average of the true shear along the line of sight,
\begin{equation}
    \bar{g}^{\mathrm{obs}}\left( \pmb{\theta} \right) = \int_{0}^{\infty} n_\gamma (z) g^{\mathrm{true}}\left(\pmb{\theta}, z \right) \d{z},
\end{equation}
where the weight $n_\gamma (z)$ is the \textit{effective redshift distribution} for lensing. When the redshift $z_j$ and shear response $R_j$ are both known for each galaxy $j$, the weight can be assessed as
\begin{equation}
    n_\gamma (z) = \frac{\Sigma_j R_j \delta(z-z_j)}{\Sigma_j R_j}.
    \label{eqn:weight}
\end{equation}
In other words, each galaxy is an individual sample of the underlying shear field. In the presence of unrecognized blends, however, equation~(\ref{eqn:weight}) no longer holds, since neither $z_j$ nor $R_j$ are accurate for the blended sources. \citet{des_bl} have shown that, in addition to shear-dependent selection as introduced in \citet{selection_shear}, a blended source also contributes to the shear bias by responding to all of its components at different redshifts. This effect is referred to as the ``shear interplay'' in \citet{kids_shear}, which concluded that this  effect is negligible for KiDS, but not for future deeper weak lensing surveys. To account for this issue (as well as redshift-dependent shear response) in DES Y3, \citet{des_bl} chose to calibrate directly the relation between $n_\gamma(z)$ and $n(z)$ with an empirical model fit to image simulation, where $n(z)$ is the observed redshift distribution of sources.

We have developed a method to identify likely blends using only catalog information from ground-based imaging. Any analysis that could be biased by including unrecognized blends could utilize this method to reduce (though not eliminate) the number of blends in the sample. To date, only a few studies have quantified the impact of unrecognized blends on cosmology. \citet{erfan} studied the effects of blending on cosmic shear, based on the {\tt Buzzard} synthetic galaxy catalog \citep{buzzard} of 10254 square degrees and an equivalent depth of 1.3 years of LSST. A $2\sigma$ decrease in the derived structure growth parameter $S_8 = \sigma_8 (\Omega_{\mathrm{m}}/0.3)^{0.5}$ is observed due to unrecognized blends, which is comparable to the ``tension'' between the DES Y3 cosmic shear measurement \citep[e.g.,][]{des_y3_cs1, des_y3_cs2} and the \textit{Planck} Cosmic Microwave Background result \citep{planck_2018}. A more recent study on galaxy clustering \citep{ben_clustering} using the simulated sky survey underlying the second data challenge (DC2) of the LSST Dark Energy Science Collaboration \citep{dc2} find a $3\sigma$ change in the two-point correlation function at small scales, but the large scales are unaffected. In future work, we will study the impact on cosmology with realistic blends rejection using methods developed in this work.

We would like to draw attention to blend detection with SOM, since SOM has been widely used for photometric redshift estimation (e.g., DES Y3: \citealp{des_som} and KiDS: \citealp{kids_som}), and will likely be used in LSST as well. Our method can then be used to detect potential blends for LSST without much extra effort.
The only input required will be a small patch of survey footprint overlapping space-based imaging to identify pure galaxies to train the SOM, and to mark ``blending cells'' in the SOM for LSST photo-\textit{z}. Then, certain proportions of the photo-\textit{z} sample can be removed or down-weighted when estimating the redshift distribution $n(z)$, yielding a cleaner measurement. In addition, SOM has been used to select spectroscopic follow-ups for photo-\textit{z} training, e.g. in the Complete Calibration of the Color-Redshift Relation \citep[C3R2;][]{dan1}. Since the spectroscopic observations are also ground-based, they suffer from the same blending issue as photometry and photo-\textit{z}. Our method can then provide extra selection on spectroscopic targets towards a cleaner spec-\textit{z} sample. This cleaner spec-\textit{z} sample can be used in other photo-\textit{z} algorithms for an improved accuracy. On the other hand, state of the art shear estimators \citep[e.g. \textsc{Metadetection}\footnote{\url{https://github.com/esheldon/metadetect}};][]{selection_shear, mdet_rubin} have evolved into the estimation of subsample-level shear response in order to mitigate contamination of single-object shear response $R_j$ due to blends, as mentioned above. As an example, KiDS used a shear calibration method similar to a \textsc{Metadetection} pipeline \citep[referred to as ``self-calibration'' in][]{kids_self_cal}, where the shear bias is calculated within bins based on galaxy size, PSF size, and SNR. 
SOM can be used for such binning schemes as well as for more complicated ones, and can work with \textsc{Metadetection} to provide cell-based shear response for LSST. Thus, SOM has the potential to bridge photometric redshift estimates and shear calibration, and provides a mechanism for blending studies on both sides.

%% file: summary.tex
\section{Summary}
\label{sec:sum}

In this work, we explore the use of several Machine Learning methods to detect unrecognized blends using only catalog-level information. We use object colors and sizes as training features, since blended objects can have colors rarely observed in a normal galaxy population, and they are expected to be larger and brighter on average than non-blended objects. 

The Machine Learning algorithms that we utilize include Self-Organizing Maps (SOM), Random Forest (RF), k Nearest Neighbors (k-NN), and a few anomaly detection algorithms.  We test the methods on the ground-based COSMOS data set \citep{laigle}, where high-resolution HST imaging is available \citep{hst1, hst2} as the truth for blends.
Using a 2-step matching algorithm on the COSMOS dataset, we find that 83\,\% of objects are classified as ``pure'' objects and 17\,\% of objects are unrecognized blends at $i^{\p}<24.5$.



We quantify the performance of each algorithm with the fraction of successfully identified unrecognized blends (Recall) and the fraction of all detections labeled as blends (Cost). We find a similar performance between the SOM blending cell method and k-NN, while RF slightly outperforms the two and SOM distance cut is significantly less efficient.
In general, the best-performing algorithms can identify approximately 30\,\% / 80\,\% of unrecognized blends at the cost of 10\,\% / 50\,\% of all detected sources, respectively.

We then study a variety of modifications to the algorithms to better characterize our ability to classify unrecognized blends.
Using a more stringent definition of unrecognized blends (``strong blends'' only) provides better recall.
SOM and RF achieve similar results regardless of algorithm-specific choices, e.g., SOM distance metrics or RF classification vs regression.
For the upcoming LSST, we find that the \textit{i}-mag and size features are some of the most important to include while the NIR bands (of the same spatial resolution) can be dropped with minimal impact on ability to detect unrecognized blends.

Parallel to detecting unrecognized blends, we investigate how removing unrecognized blends can help to reduce photo-\textit{z} outliers.
We assemble a spectroscopic redshift catalog from several sources to assess the quality of photometric redshift estimates in the dataset in order to identify photo-\textit{z} outliers.
While there are 30-band photo-\textit{z} estimates in the COSMOS dataset, we train our own SOM\textit{z} estimator to better mimic LSST conditions.
We find that the photo-\textit{z} outliers generally have more exotic colors than unrecognized blends, and algorithms aimed at ``feature outliers'' (SOM distance cut and LOF) are the better options for detecting them. This gives SOM several synergies as the same SOM can be used for multiple purposes: labeling unrecognized blends through blend-cell occupancy ratio, photo-\textit{z} outlier through distance cut, and a mechanism to study the impact of blends on photo-\textit{z} and shear measurements.

This work has potential application in Rubin-LSST, where 20\%-30\% of detections are expected to be unrecognized blends \citep{roman_synth}.
A genuine solution to unrecognized blends could be to match LSST with space-based imaging, like Euclid, to label such objects.
The methods developed in this work offer a possible mitigation to the unrecognized blends problem in LSST before Euclid / Roman imaging are available for the entire LSST area, as well as the possibility to train an unrecognized blend detector on the overlapping area and apply to the entire area. Thus, the methodology developed here can lead to a cleaner galaxy sample and potentially reduce systematic uncertainties in LSST cosmology caused by unrecognized blends.

%% file: ack.tex
\section*{Author Contributions}
SL and AvdL conceived and initiated the project. SL created the matched catalog and SOM implementations. PA created the RF, k-NN, and anomaly detection implementations. SL and PA wrote the paper and created figures. All authors discussed the results and implications. AvdL gave detailed guidance on the study and mentored SL and PA along with commenting on the manuscript at all stages.

\section*{Acknowledgements}

The authors thank Ricardo Herbonnet, Remy Joseph, Patricia Burchat, and Ben Levine for illuminating discussion and suggestions. 

This paper has undergone internal review in the LSST Dark Energy Science Collaboration. We are grateful to our internal reviewers, Patricia Burchat and Tianqing Zhang, for their thoughtful feedback.

We also thank the anonymous referees for helpful comments and suggestions to this paper.

SL and AvdL were supported by the U.S. Department of Energy under award DE-SC0018053. AvdL and PA are supported by the U.S. Department of Energy under awards DE-SC0025309 and DE-SC0023387. PA was also supported in part by the Stony Brook Lourie Fellowship. SL is also supported in part by the U.S. Department of Energy under grant number DE-1161130-116-SDDTA and under Contract No. DE-AC02-76SF00515 with the SLAC National Accelerator Laboratory.

The DESC acknowledges ongoing support from the Institut National de 
Physique Nucl\'eaire et de Physique des Particules in France; the 
Science \& Technology Facilities Council in the United Kingdom; and the
Department of Energy, the National Science Foundation, and the LSST 
Corporation in the United States.  DESC uses resources of the IN2P3 
Computing Center (CC-IN2P3--Lyon/Villeurbanne - France) funded by the 
Centre National de la Recherche Scientifique; the National Energy 
Research Scientific Computing Center, a DOE Office of Science User 
Facility supported by the Office of Science of the U.S.\ Department of
Energy under Contract No.\ DE-AC02-05CH11231; STFC DiRAC HPC Facilities, 
funded by UK BEIS National E-infrastructure capital grants; and the UK 
particle physics grid, supported by the GridPP Collaboration.  This 
work was performed in part under DOE Contract DE-AC02-76SF00515.

This work used \textsc{randomForest} \citep{randomForestR}, \textsc{class} \citep{S_class} and \textsc{scikit.learn} \citep{scikit-learn} for Random Forest and k-Nearest Neighbors studies. Some general-purpose packages used in this research include: Astropy, a community-developed core python package for Astronomy \citep{astropy}; \textsc{TOPCAT} \citep{topcat}, a graphical user tool for interactive table manipulation; \textsc{STILTS} \citep{stilts}, a command-line based tool for the processing of tabular data; \textsc{healpy}, a HEALPix package \citep{hp1, hp2}; \textsc{Numpy} \citep{np}, \textsc{Matplotlib} \citep{plt}, \textsc{ggplot2} \citep{ggplot}, \verb|randomForest| \cite{randomForestR}. We adopt color-blind friendly schemes from www.ColorBrewer.org by Cynthia A. Brewer, Geography, Pennsylvania State University.

This research has made use of the NASA/IPAC Infrared Science Archive, which is funded by the National Aeronautics and Space Administration and operated by the California Institute of Technology.

This research has made use of the VizieR catalogue access tool, CDS, Strasbourg, France (DOI: 10.26093/cds/vizier). The original description of the VizieR service was published in 2000, A\&AS 143, 23.

This research used data obtained with the Dark Energy Spectroscopic Instrument (DESI). DESI construction and operations is managed by the Lawrence Berkeley National Laboratory. This material is based upon work supported by the U.S. Department of Energy, Office of Science, Office of High-Energy Physics, under Contract No. DE–AC02–05CH11231, and by the National Energy Research Scientific Computing Center, a DOE Office of Science User Facility under the same contract. Additional support for DESI was provided by the U.S. National Science Foundation (NSF), Division of Astronomical Sciences under Contract No. AST-0950945 to the NSF’s National Optical-Infrared Astronomy Research Laboratory; the Science and Technology Facilities Council of the United Kingdom; the Gordon and Betty Moore Foundation; the Heising-Simons Foundation; the French Alternative Energies and Atomic Energy Commission (CEA); the National Council of Science and Technology of Mexico (CONACYT); the Ministry of Science and Innovation of Spain (MICINN), and by the DESI Member Institutions: www.desi.lbl.gov/collaborating-institutions. The DESI collaboration is honored to be permitted to conduct scientific research on Iolkam Du’ag (Kitt Peak), a mountain with particular significance to the Tohono O’odham Nation. Any opinions, findings, and conclusions or recommendations expressed in this material are those of the author(s) and do not necessarily reflect the views of the U.S. National Science Foundation, the U.S. Department of Energy, or any of the listed funding agencies.

%% file: data.tex
\section*{Data Availability}

The data sets underlying this article are publicly available. They can be accessed at:
\begin{enumerate}
    \item[] COSMOS 2015: The dataset is available at IRSA under the DOI number \doi{10.26131/IRSA527}, and is available on the COSMOS website \url{https://cosmos.astro.caltech.edu/page/photom}
    \item[] HST-COSMOS: The dataset is available at IRSA under the DOI number \doi{10.26131/IRSA169}, and is available on the COSMOS website at \url{https://cosmos.astro.caltech.edu/page/hst}
    \item[] zCOSMOS: from the zCOSMOS website at \url{http://cesam.lam.fr/zCosmos/}
    \item[] DEIMOS: through the VizieR service at \url{https://cdsarc.cds.unistra.fr/viz-bin/cat/J/ApJ/858/77}
    \item[] C3R2: from the C3R2 website at \url{https://sites.google.com/view/c3r2-survey/home?authuser=0}
    \item[] VUDS: from the VUDS project website at \url{https://cesam.lam.fr/vuds/}
    \item[] DESI EDR: from the DESI project website at \url{https://data.desi.lbl.gov/doc/releases/}
\end{enumerate}

%% file: app.tex
\section{Detailed Data Selection}
\subsection{HST Data}
The selections applied to the HST data include:
\label{sec:sel_hst}
\begin{enumerate}[leftmargin=*,labelindent=10pt,label=\arabic*.]
    \item  {\tt 0 < mag\_best < 26.5}. \\
    We keep sources that are up to 2 mags fainter than the base COSMOS sample ({\tt ip\_MAG\_AUTO < 24.5}) to allow for blending with fainter sources. 
    \item  {\tt 0 < flux\_radius < 10/0.049} \\
    We remove sources with negative flux radii, as well as diffuse sources larger than 10 arcsecs. The large diffuse sources would contaminate the matching process for blends if not removed. The flux radius is in units of pixels, and the HST ACS pixel scale\footnote{\url{https://hst-docs.stsci.edu/acsdhb/chapter-1-acs-overview/1-1-instrument-design-and-capabilities}} is 0.049 arcsec/pix.
    \item  {\tt mu\_class = 1 or 2}. \\
    The {\tt mu\_class} parameter is a morphology-based star-galaxy separation flag derived from the {\tt MU\_MAX} and {\tt MAG\_AUTO} parameters from {\tt SExtractor}. We keep galaxies ({\tt mu\_class = 1}) and point sources ({\tt mu\_class = 2}), and remove ``fake'' detections with {\tt mu\_class = 3}. 
\end{enumerate} 

\subsection{Ground-Based COSMOS Data}
\label{sec:sel_ground}
We apply the following selections on the ground-based COSMOS catalog \textbf{after} matching to HST:
\begin{enumerate}[leftmargin=*,labelindent=10pt,label=\arabic*.]
    \item {\tt FLAG\_COSMOS=1} \& {\tt FLAG\_HJMCC=0} \& {\tt FLAG\_PETER=0}. \\
        Sources that satisfy these cuts are kept. The first two flags mark sources within the COSMOS area 
        that also have UltraVISTA coverage. The last flag removes saturated sources and masked areas in optical bands.
    \item {\tt b\_FLAGS < 4}. \\
        Photometry flags from {\tt SExtractor}. We keep sources with {\tt b\_FLAGS = 0, 1 and 2}, which corresponds to isolated sources, sources with bright neighbors, and recognized blends, respectively. Here {\tt b} stands for all of the \textit{uBVr$i^{\p}z^{\pp}$YJH} bands; sources that satisfy this cut in all bands are kept.
    \item {\tt b\_IMAFLAGS\_ISO = 0}. \\
        Only unsaturated sources in all bands are kept.
    \item {\tt b\_FLUXERR} > -99. \\
        We remove non-observations in any bands. There are only 8 of them.
    \item {\tt FLUX\_RADIUS > 0}. \\
        We remove sources with negative flux radii.
    \item {\tt ip\_FLUX\_APER3/FLUX\_RADIUS$^2$ < 0.002}. \\
        Sources with extremely low surface brightness are removed.
    \item {\tt TYPE = 0}. \\
    \edit{Star / galaxy / AGN separation based on LePHARE SED fitting and X-ray detection. Only non-AGN galaxies are kept.}
    \item {\tt ip\_MAG\_AUTO < 24.5}. \\
    This magnitude cut at $i^{\p}<24.5$ is limited by the depth of the ``truth'' catalog -- the HST COSMOS catalog. We allow galaxies to blend with sources up to 2 magnitude fainter, and the HST COSMOS catalog is complete at $\sim 26.5$.
    \item {\tt ZPDF > 0}. \\
    The COSMOS 30 band photo-z estimated with galaxy SED templates, measured at the median of the likelihood distributions. Only sources with positive photo-z measurement are used for training.
\end{enumerate}

\subsection{Spectroscopy Data}
\label{sec:sel_spec}
We assemble a spectroscopy sample from the following catalogs:
\begin{enumerate}[leftmargin=*,labelindent=10pt,label=\arabic*.]
    \item zCOSMOS \citep{zcosmos1, zcosmos2}: zCOSMOS is a large survey in the COSMOS field conducted with the VIMOS spectrograph on the Very Large Telescope (VLT). The survey is designed to characterize the galaxy environments and produce diagnostic information on the galaxies and active galactic nuclei. We download the zCOSMOS DR3 catalog which contains 20689 objects, and apply cuts to select high quality redshift measurements. Following the data release document\footnote{\url{https://irsa.ipac.caltech.edu/data/COSMOS/spectra/z-cosmos/zCOSMOS_DR3.pdf}}, we keep the Confidence Class (cc) flag of: 
    \begin{itemize}
        \item 4.x and 3.x: (very) secure redshift measurements.
        \item 9.5: one-line redshifts that are consistent with photometric redshifts at 
            $|z_{\mathrm{spec}} - z_{\mathrm{phot}}|/(1+z_{\mathrm{spec}}) < 0.04$. 
        \item 2.5: Probable redshifts that are consistent with photometric redshifts at 
            $|z_{\mathrm{spec}} - z_{\mathrm{phot}}|/(1+z_{\mathrm{spec}}) < 0.04$.
    \end{itemize}
    The selected sample contains 15867 sources.
    \item DEIMOS \citep{deimos1, deimos2}: The COSMOS DEIMOS Catalog consists of 10718 objects in the COSMOS field, observed through multi-slit spectroscopy with the Deep Imaging Multi-Object Spectrograph (DEIMOS) on the Keck II telescope. We keep sources with a quality flag {\tt Q} = 2 or 1.5, representing reliable spectroscopic identification or sources with 
    $|z_{\mathrm{spec}} - z_{\mathrm{phot}}|/(1+z_{\mathrm{spec}}) < 0.1$, respectively.
    We also remove sources with $z_{\mathrm{spec}} > 10$. This results in 8339 sources from DEIMOS.
    \item C3R2 \citep{dan1, dan2, stanford, lbt}: The Complete Calibration of the Color-Redshift Relation (C3R2) is a spectroscopic survey at depth $i\sim24.5$. C3R2 aims to fill out the galaxy color space with spectroscopic redshifts, to provide a firm foundation for photometric-redshift calibration for upcoming weak lensing cosmology surveys. The C3R2 catalogs (DR1+DR2+DR3+LBT) consist of $\sim5000$ specstropic redshift measurements from Keck, VLT and LBT, of which 2930 sources are within the COSMOS region. We keep 2501 of those with {\tt flag = 4}, which have highly certain redshift estimates based on multiple line detections.
    \item VUDS \citep{vuds1, vuds2}: The VIMOS Ultra Deep Survey (VUDS) is a spectroscopic redshift survey of $\sim10000$ very faint galaxies to study the major phase of galaxy assembly. VUDS covers 3 separate fields: COSMOS, ECDFS and VVDS-02h, providing an additional 384 sources in the COSMOS field. We keep only 144 sources with {\tt zflags} = 3 or 4, which have moderate to high S/N with several absorption and/or emission lines, and strong cross–correlation signal with good to excellent continuum match to templates.
    \item DESI EDR \citep{desi_edr}: The Early Data Release of the Dark Energy Spectroscopic Instrument (DESI) contains 1.2 million extra-galactic sources (galaxies and quasars), which makes up 2\% of the final data set. We keep sources with {\tt NPIXELS>0} and {\tt ZWARN=0} and remove stars ({\tt SPECTYPE=`STAR'}). This yields a sample of 10509 sources within the COSMOS field.
\end{enumerate}

\section{Calculating Luptitudes}
\label{sec:appb}
\begin{figure}
    \centering
    \includegraphics[width=\columnwidth]{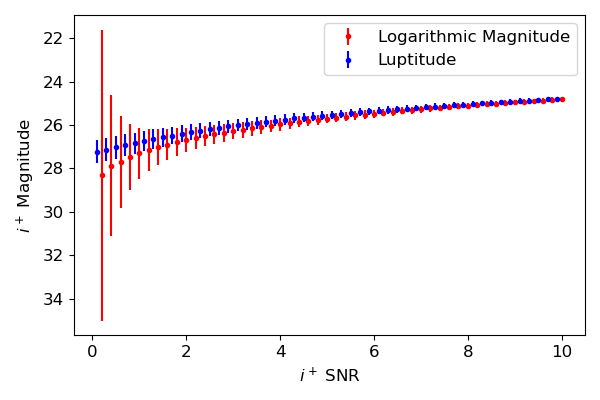}
    \caption{Comparison of Logarithmic magnitude and asinh Luptitude in the COSMOS $i^{\p}$ band. The magnitude uncertainties are calculated from median fluxes and flux errors in each S.N.R bin. Note that there is no difference between the two magnitudes at $i^{\p}<24.5$.}
    \label{fig:luptitudes}
\end{figure}

The COSMOS photometry includes non-detections, i.e. entries with zero or negative flux in some bands which causes problems in magnitudes and colors. Instead of removing these non-detections, which would introduce unwanted selection effects as explained in Sect.~\ref{subsec:match}, we adopt asinh magnitudes for the base COSMOS sample following \citet{lupton}. They are calculated with 
\begin{equation}
    m = -a \left[ \sinh^{-1}{\left( \frac{f/f_0}{2b} \right) }+ \ln{b} \right],
\end{equation}
where $a=2.5\log_{10}{e} $ is a constant, and $f$ is the flux of each source normalized by $f_0 = 3631 \times 10^6 \mathrm{\mu Jy}$, the reference flux of an object at magnitude 0. The ``softening parameter'' $b$ sets the small scale at which the magnitude responds linearly to the normalized flux. \citet{lupton} suggest setting $b$ to be the normalized flux of an object with a signal-to-noise ratio of about one. In our practice, we set $b$ to be the median normalized flux of sources with $0.9 < \mathrm{S.N.R} < 1.1 $, which are listed in Table~\ref{tab:soft}. The magnitude uncertainties are calculated with:
\begin{equation}
    m_{\rm err} = \frac{a \cdot f_{\rm err}/f_0}{\sqrt{4b^2 + (f/f_0)^2}}.
\end{equation}
Luptitudes are designed to converge into logarithmic magnitudes at higher S.N.R, and have meaningful uncertainties at lower or even negative flux level due to its asymptotic linear relation with flux. A comparison between the $i^{\p}$ band Logrithmic magnitudes ({\tt ip\_MAG\_AUTO}) from COSMOS and the calculated Luptitudes can be found in Fig.~\ref{fig:luptitudes}. 

\begin{table}
	\label{tab:soft}
	\centering
	\begin{tabular}{c c c c}
		\hline
		           &                 & Zero-Flux Magnitude &           \\
	    Filter     & $b$[$10^{-11}f_0$] & $m(f/f_0=0)$ & $m(f/f_0=10b)$   \\ \hline
        \textit{u} &      1.12          &     27.39     &   24.88         \\
        \textit{B} &      0.45          &     28.38     &   25.87         \\
	    \textit{V} &      0.85          &     27.67     &   25.16         \\
		\textit{r} &      0.77          &     27.79     &   25.28         \\
		$i^{\p}$   &      1.19          &     27.31     &   24.80         \\
		$z^{\pp}$  &      2.10          &     26.70     &   24.19         \\
		\textit{Y} &      3.85          &     26.03     &   23.52         \\
	    \textit{J} &      4.17          &     25.95     &   23.44         \\
		\textit{H} &      5.93          &     25.57     &   23.06         \\
        \hline
    \end{tabular}
    \caption{Softening parameters to calculate the asinh magnitudes for the COSMOS data set.}
\end{table}

\section{Anomaly Detection}
\label{sec:anom}
\begin{figure}
     \centering
     \includegraphics[width=\linewidth]{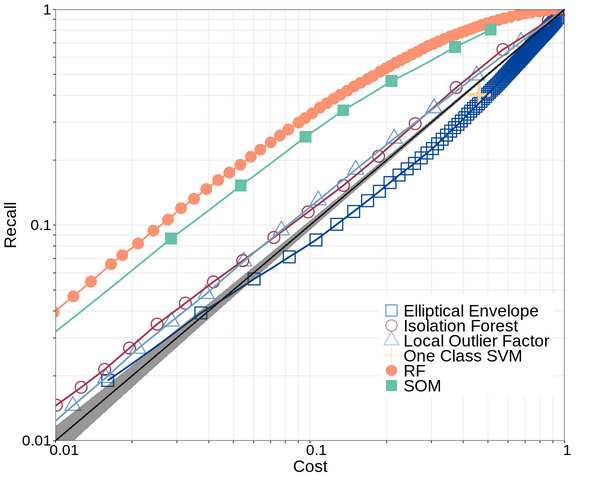}
     \caption{Comparison of RF and SOM against a variety of anomaly detection methods. }
     \label{fig:anomaly_full}
 \end{figure}

Anomaly detection algorithms seem like promising methods as we expect (some) blended galaxies to have different colors when compared to pure galaxies i.e are outliers. 
We investigate 4 popular anomaly detection methods:
\begin{enumerate}
    \item \textbf {Elliptical Envelope} - EE is an unsupervised algorithm that identifies outliers with a Mahalanobis distance greater than some threshold \citep{EE_1, EE_2}. 
    The Mahalanobis distance is a measure of how many standard deviations a datum is from a distribution. \[
    d_{\textrm{EE}} = \sqrt{\left(\vec{x} - \vec{\mu}\right)^T C^{-1}\left(\vec{x}  - \vec{\mu}\right)}.
    \] While EE is unsupervised, we use the pure galaxies to estimate the covariance $C$ and mean $\mu$.  We can tune the performance by changing the cut-off on distance to label objects as blends.
    \item \textbf{Local Outlier Factor} - LOF is an unsupervised algorithm that detects outliers by calculating the local density compared to the nearest neighbors for a data point that was first described in \cite{breunig2000lof}. Any point that has a low density compared to its nearest neighbors is labelled as an outlier. The density in comparison to its neighbors is turned into an \textit{outlier factor} which is less than 1 for inliers. Outliers will have scores greater than 1. We can tune the performance by changing the cut-off on which outliers get marked as blends. 
    \item \textbf{Isolation Forest} - ``iForest'' is first described in \cite{liu2008isolation}. iForest is created by creating random splits between the minimum and maximum of a feature and then counting the number of splits it takes to uniquely label a datum, the \textit{path length}. iForest operates by assuming that outliers will have a shorter path length to isolate them into a terminal node with no other data or into nodes with all data sharing the same label. This is an unsupervised algorithm with no knowledge of the blend labels. We can tune this by changing the maximum path length an outlier can have.
    \item \textbf{One Class Support Vector Machine} - One Class SVM is described in \cite{svm-article}. This is an unsupervised outlier detector that attempts to create a bounded region (hypersphere) in parameter space that encloses the majority of points which are thought to be inliers. One advantage with this method is the ability to specify a kernel according to any underlying geometry however there is no tuning capability. 
 \end{enumerate}
The results are plotted in comparison to RF and SOM in Fig.~\ref{fig:anomaly_full}.
As noted above, One Class SVM has no tunability so it is shown as a singular point on the recall-cost plot.
Overall the anomaly detection methods are unable to catch unrecognized blends. LOF and iForest do the best while EE is worse than removing points at random.

\section{Classification Metrics}
\label{sec:app-class}
\begin{figure}
    \centering
    \includegraphics[width=0.9\linewidth]{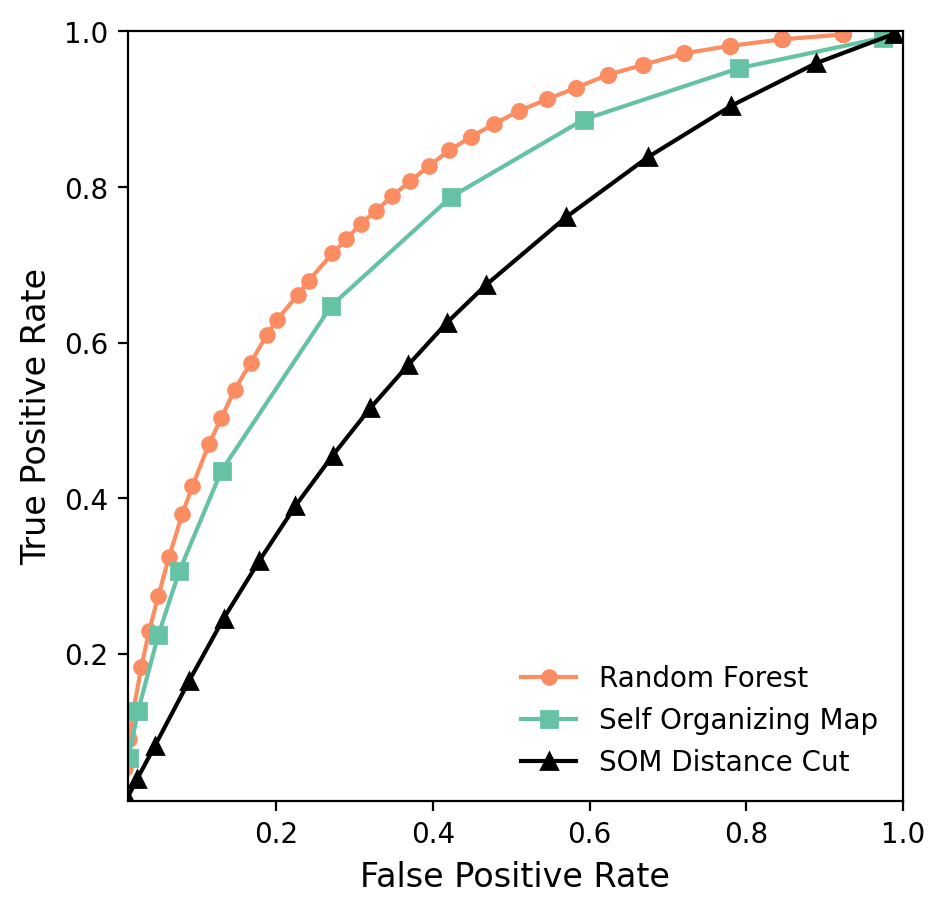}
    \caption{\edit{Receiver Operating Characteristic (ROC) curve for a few methods developed in the paper with RF in orange circles, blend ratio in green squares, and distance cut in black triangles.}}
    \label{fig:som_rf_roc}
\end{figure}

\begin{figure*}
    \centering
    \includegraphics[width=.9\linewidth]{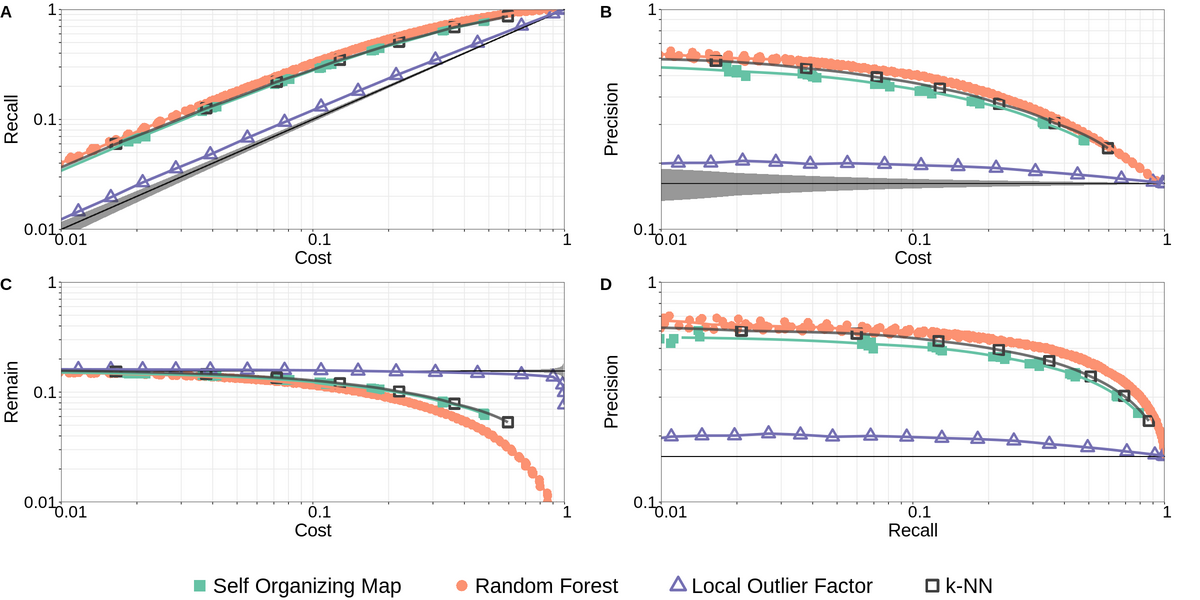}
    \caption{Removing potential blended sources from the validation sample using different algorithms and quantifying the performance in terms of Cost, Recall, Precision, and Remaining Fraction  all of which are defined in Eq.~\ref{eq:precision_recall}. The SOM (blending cell) results are shown in green squares, RF in orange circles, LOF in purple triangles, and k-NN in hollow black squares. We include a baseline curve in gray representing a random, or no-skill, classifier along with a 95\% CI derived from a hypergeometric distribution as the shaded region around the baseline. One anomaly detection method, Local Outlier Factor, is plotted as a comparison. It has the best results out of the 4 anomaly detection methods considered in this work, yet it behaves marginally better than a random classification, while RF, SOM, and k-NN have substantial better performances. }
    \label{fig:classification}
\end{figure*}

Metrics that are relevant for classification include \textbf{Cost}, \textbf{Recall}, \textbf{Precision}, \edit{\textbf{False Positive Rate (FPR)}} and \textbf{Remain}, which are defined as follows. 
\edit{We use the same notation as Eq.~\ref{eq:recall_cost} which we repeat here for clarity. Note that \textbf{Recall} is equivalent to the \textbf{True Positive Rate} (TPR)}.
Let $B$ be the total number of unrecognized blends in the validation sample and $P$ the total number of pure galaxies. Then $B+P$ is the validation sample size. Let $R_b$ be the number of removed unrecognized blends, and $R_p$ the number of removed pure galaxies, then $R_b + R_p$ is the removed sample size, and $B + P - R_b - R_p$ is the remaining sample size after the removal. We define
\begin{equation}
\begin{split}\label{eq:precision_recall}
    &\textrm{Cost} = \frac{R_b+R_p}{B+P}; \;\;\; \textrm{Recall} = \frac{R_b}{B} = \textrm{TPR}; \;\;\textrm{FPR} = \frac{R_p}{P}; \\
    &\textrm{Precision} = \frac{R_b}{R_b+R_P}; \;\;\;\; \textrm{Remain} = \frac{B-R_b}{B+P-R_b-R_p}.
\end{split}    
\end{equation}
In words, Cost is the fraction of sample being removed; Recall is the fraction of total blends being removed; Precision is the fraction of blends among the removed sample; and Remain is the fraction of blends after removing certain objects. Ideally we would have high Recall, high Precision, and low Remain, all at a low Cost. 
In Fig.~\ref{fig:classification}, we repeat Fig~\ref{fig:res} as Panel A, and include 3 other ways to understand the performance of the classifiers: (B) Remain vs Cost, (C) Precision vs Cost, and (D) Precision vs Recall. 
\edit{A typical Receiver Operating Characteristic (ROC) curve for binary classification (TPR versus FPR) is shown in Fig~\ref{fig:som_rf_roc}.
While a useful tool for judging binary classification in general, we found that replacing the FPR with the cost allows us to better understand the change in total catalog size.}

We also show related statistics in Table.~\ref{tab:fullres}.


\begin{table*}
\caption{Summary statistics of the SOM and RF blend detection. The left half shows the results for removing blending SOM cells and the right half is removing high-score blending objects as predicted by Random Forest. Columns are defined in equation~(\ref{eq:precision_recall}). The results shown are averaged over 5 realizations. All numbers are in percentiles except for the Cell number (first column). Results from the 5 realizations are plotted in Fig.~\ref{fig:classification}.}
\centering
\begin{tabular}{RRRRRRRRR}
\toprule
\multicolumn{5}{c}{SOM: Remove Blending Cells}  &  \multicolumn{4}{c}{RF: Remove Predicted Blends} \\
\cmidrule(lr){1-5}
\cmidrule(lr){6-9}

\multicolumn{1}{r}{Cell}       &
\multicolumn{1}{r}{Cost}       &
\multicolumn{1}{r}{Recall}     &
\multicolumn{1}{r}{Precision}  &
\multicolumn{1}{r}{Remain}     &
\multicolumn{1}{r}{Cost}       &
\multicolumn{1}{r}{Recall}     & 
\multicolumn{1}{r}{Precision}  &
\multicolumn{1}{r}{Remain}     \\ \midrule
 10   & 0.77  & 2.64   &   53.46   & 15.20  & 0.80  &  3.29  &   62.71   & 15.16  \\
 50   & 3.16  & 9.97   &   48.88   & 14.40  & 3.19  & 11.98  &   59.82   & 14.13   \\
 100  & 5.90  & 17.21  &   45.17   & 13.63  &  5.45 & 19.22  &   56.94   & 12.80  \\
 200  & 10.58 & 28.02  &   41.03   & 12.47  & 10.60 & 33.63  &   51.30   & 11.48  \\
 300  & 14.61 & 36.14  &   38.33   & 11.58  & 14.67 & 42.98  &   47.32   & 10.33  \\
 500  & 22.01 & 48.58  &   34.20   & 10.21  & 22.71 & 57.83  &   41.84   & 8.39   \\
 1000 & 38.20 & 68.56  &   27.80   &  7.88  & 38.47 & 77.14  &   32.41   & 5.66   \\
 1500 & 53.16 & 80.96  &   23.59   &  6.30  & 54.57 & 89.35  &   26.46   & 3.61   \\
 \bottomrule
\end{tabular}
\label{tab:fullres}
\end{table*}

\section{Feature Variants}
\label{sec:feat_varia}
Here we show the results for more RF and SOM setups in addition to Sect.~\ref{subsec:alt_conf}.

Alternative Configurations for SOM:
\begin{enumerate}[leftmargin=*,labelindent=10pt,label=\arabic*.]
\setcounter{enumi}{6}
    \item 10+10, Euclidean: using Euclidean Distance instead of $\chi^2$ distance (equation~[\ref{eqn:dist}]) for training the SOM and for mapping objects onto the SOM.
    \item 10+10, Blends: using both the pure training sample and the identification sample to train the SOM. 
    \item 10+10, Counts: Removing cells based on the counts of blends in the cell instead of the ratio of blends to pure galaxies in the cell.
\end{enumerate}

Alternative Configurations for RF:
\begin{enumerate}[leftmargin=*,labelindent=10pt,label=\arabic*.]
\setcounter{enumi}{9}
    \item 10+10, Regression: using 10 features for training and testing and treating the labels as integers (0 for pure and 1 for blends). Create a regression forest that outputs a score between 0 to 1. The main text uses a classification forest which is then turned into a score by counting the number of trees voting for ``pure'' as outlined in Sect.~\ref{subsec:RF_config}. This method directly estimates a score for each datum. 
    \item 10+10, 2-class: using 10 features for training and testing on a classification forest with the labels ``pure'' and ``blend.'' The main text gives three labels: ``pure'', ``weak'', ``strong.''
    \item 19+19, training with photometric uncertainties: using 19 features for training and testing with the features being 8 colors, 8 color errors, 1 magnitude, 1 magnitude error, and flux radius. 
\end{enumerate}

\begin{figure*}
	\centering
        \includegraphics[width=.9\textwidth]{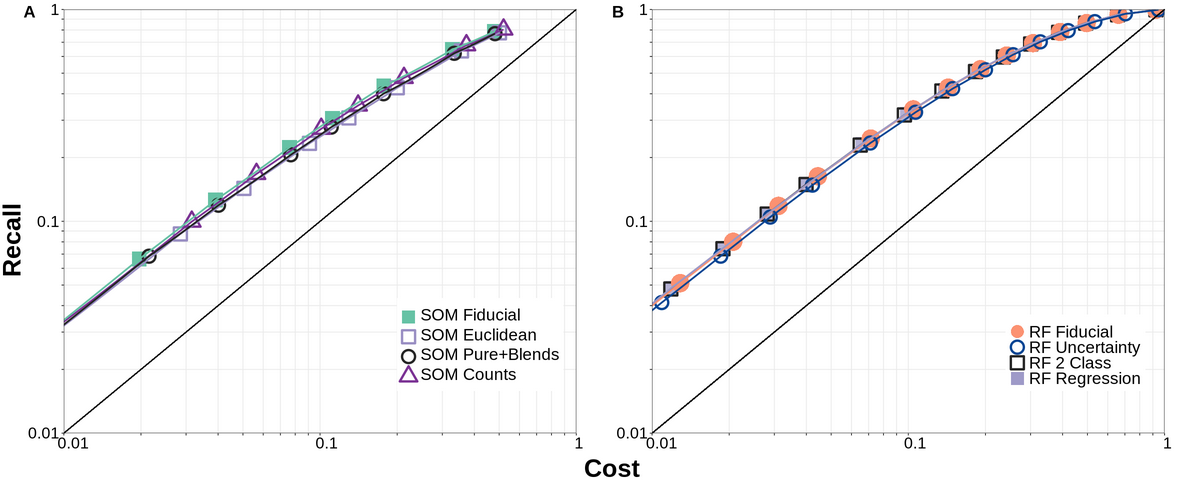}
	\caption{Alternative configurations for SOM (left) and RF (right). In all cases, we see no apparent changes in performance.} 
	\label{fig:alt_conf}
\end{figure*}

These alternative configurations are displayed in Fig.~\ref{fig:alt_conf}. We found no evident change in performance in any cases.
